\documentclass[journal]{IEEEtran}
\usepackage{bm}
\usepackage{graphicx}
\usepackage{amsmath}
\usepackage{indentfirst}
\usepackage{bm}
\usepackage{enumerate}

\usepackage{multirow} 
\usepackage{cite}
\usepackage{array}
\usepackage{amsmath,amssymb,amsfonts}
\usepackage{algorithm,algorithmic}
\usepackage{graphics}
\usepackage{epsfig}
\usepackage[tight,footnotesize]{subfigure}
\usepackage{enumerate}
\usepackage{pifont}
\usepackage[english]{babel}
\usepackage{xparse}				
\usepackage{multirow, booktabs} 
\usepackage[table]{xcolor}		
\usepackage{verbatim}
\usepackage{mathrsfs}

\ifCLASSINFOpdf
\else
\fi
\usepackage{amsmath}
\usepackage{diagbox}

\begin{document}
%
\title{Multi-Tones' Phase Coding (MTPC) of Interaural Time Difference by Spiking Neural Networks}
%
%
%

\author{Zihan~Pan, Malu~Zhang, Jibin Wu, 
        and~Haizhou~Li,~\IEEEmembership{Fellow,~IEEE}

\thanks{Zihan Pan and Malu Zhang contributed equally in this work, and should be regarded as co-first authors.}
\thanks{Jibin Wu is the corresponding author.}
}

%
%

\markboth{Journal of \LaTeX\ Class Files,~Vol.~14, No.~8, August~2015}%
{Shell \MakeLowercase{\textit{et al.}}: Bare Demo of IEEEtran.cls for IEEE Journals}
%



\maketitle

\begin{abstract}
Inspired by the mammal's auditory localization pathway, in this paper we propose a pure spiking neural network (SNN) based computational model for precised sound localization in the noisy real-world environment, and implement this algorithm in a real-time robotic system with microphone array. The key of this model relies on the MTPC scheme, which encodes the interaural time difference (ITD) cues into spike patterns. This scheme naturally follows the functional structures of human auditory localization system, rather than artificially computing of time difference of arrival. Besides, it highlights the advantages of SNN, such as event-driven and power efficiency. The MTPC is pipeliend with two different SNN architectures, the convolutional SNN and recurrent SNN, by which it shows the  applicability to various SNNs. This proposal is evaluated by the microphone collected location-dependent acoustic data, in a real-world environment with noise, obstruction, reflection or other affects. The experiment results show a mean error azimuth of $1^\circ \sim 3^\circ$, which surpasses the accuracy of the other biologically plausible neuromorphic approach for sound source localization.

\end{abstract}


%
\IEEEpeerreviewmaketitle

\section{Introduction}
\label{sec: introduction}

The ability to detect the source of sounds is one of the most crucial skills of surviving in dynamic environments for humans or other mammals \cite{brown2005comparative}. It helps to locate the prey, escape from the predators, find mates, or perform other activities. The location-dependent sounds or acoustic stimulus naturally contain redundant localization information and are sensed by the ears, encoded into neuronal spiking form of cues, and decoded into azimuths along the auditory pathways, including the auditory peripheral system and mid-brain cortex \cite{grothe2010mechanisms}.

Date back to one century ago, people started the research of auditory perception of space \cite{thompson1882li}\cite{strutt1907our}, from then on we know that the mammalian auditory localization systems benefit from two major cues: the interaural time difference (ITD) and the interaural intensity difference (IID), which are known as ``duplex theory of sound localization''. Assuming a sound source locating at the left side, the sound emitted by it will arrive at the left ear before the right ear, as well as stimulates stronger neuronal responses in the left ear. Such a propagation time delay difference or sound pressure level decaying difference can provide the information for locating the azimuth of the sound source at some spatial resolution. The ITD and IID cues are extracted in the medial superior olive (MSO) \cite{yin1990interaural} and lateral superior olive (LSO) \cite{tollin2003lateral}, respectively. They have different working frequencies, for example, low frequencies (in humans 20Hz to 2kHz) for MSO and high frequencies (in humans $>$2kHz) for LSO \cite{wall2012spiking}. From the information flow point of view, the location-dependent information from the sounds, or aforementioned time or intensity differences, are encoded into spiking pulses in the MSO and LSO. After that, they are projected to the inferior colliculus (IC) in the mid-brain, where the IC integrates both cues for estimating the sound source direction \cite{yin2002neural}. The most successful model for ITD encoding in MSO and cortex decoding might be the ``place theory'' by an American psychologist named Lloyd Jeffress \cite{jeffress1948place}, which is also referred to as Jeffress model hereafter.

In this theory, there are a series of nerve fibers with different synaptic delays used as ``delay lines'' and a coincidence detector used to detect the synchronization between binaural inputs. A distribution of spiking rates indicating the azimuth, so-called ``place map'' is generated from the synchronizing detection. By the processing, the binaural acoustic inputs are encoded into a spike pattern with ITD cues embedded. Such theory is biologically evaluated by the evidences of the ``place map'' or ``ITD map'' later found in barn owl \cite{carr1988axonal}\cite{carr1990circuit}. 
For the LSO encoding intensity difference, various models are also proposed. The LSO has a much smaller neuronal scale compared with the MSO (10,000-11,000 vs 2,500-4,000 neurons, \cite{moore2000organization}). Furthermore, the ITD cue encoded by MSO achieves a localization resolution as high as $1^\circ \sim 2^\circ$ \cite{lewicki2006sound}, which is sufficient for real-world application.
Therefore, for our computational model in this paper, we will extend the original Jeffress MSO model and propose a novel ITD-based encoding/decoding scheme for the real-world robust sound localization, together with a spiking neural network. 

Above all, those aforementioned theories or models, from the encoding pathways to the decoding mid-brain cortex, are biologically plausible, which are based on the temporal-related spiking neuronal systems. On the other hand, from the engineering application point of view, the spiking neuronal models can also offer additional engineering advantages, such as ultra-low power consumption and high-speed processing. The spiking neural networks (SNN) are believed to be the third generation of the neural network, compared with the traditional artificial neural network. Plenty of pieces of evidence and applications have proved the effectiveness and efficiency in different cognitive applications, such as computer vision, automatic speech recognition, etc. 
As such, both the biological findings on the neuronal pathway for sound localization and the SNN advantages on application motivate us to combine them together and propose a novel spiking computational model for the real-world application.

Although the neuronal circuit for encoding the binaural localization cues and neuronal processing are characterized physiologically \cite{levine1993binaural}\cite{furst2000sound}, which might be one typical case which the humans understand well about the behavioral function of a circuit in the central nervous \cite{yin2002neural}, we are still lacking good computational models that are capable of robustly solving real-world applications. Furthermore, these remarkable scientific findings can hardly tell us the realization in engineering applications, but they provide inspiration and in-depth understanding. 


Fortunately, some researchers have stepped forward to propose various SNN models for sound localization applications and positive results are provided. Experimentally derived head-related transfer function (HRTF) acoustical data from adult domestic cats were employed to train and validate the localization ability of the architecture \cite{wall2012spiking}. This work is built on the earlier work of \cite{wall2008spiking}, \cite{glackin2010spiking}. They propose a model that encodes the IID cues into the spike rates of LSO neurons and learns the spike patterns with ReSuMe rule \cite{ponulak2005resume}. It achieves $\sim 10\%$ error rates with an error tolerance of $\pm 10^{\circ}$. However, the validation sounds are single-frequency tones of 5kHz, 15kHz, and 25kHz without noise. Similar works by simulated pure tones are also reported \cite{escudero2018real}\cite{xiao2016biologically}\cite{luke2019spiking}. On the other hand, studies on ITD encoding by MSO are also presented. For example, Voutsas et al. \cite{voutsas2007biologically} propose a binaural sound source lateralization neural network (BiSoLaNN) that utilizes a single delay line to encode the ITD cues similar to the Jeffress model. This model achieves the best localization accuracy of $72.5\%$ in the range of $\pm 45^\circ$ with an error tolerance of $\pm 15^{\circ}$. However, the evaluation sounds are low-frequency pure tones between 440Hz and 1240Hz.

According to the ``duplex theory of sound localization'', some of the researchers try to combine both cues together. One of the representative studies carried out by Liu et al. \cite{liu2010biologically}, proposes a more biologically plausible SNN-based computational model, which encodes both ITD and IID cues to imitate the functions of MSO and LSO neurons, respectively. To reflect the biological organization, the MSO and LSO have separated working frequency bands ( $<$1kHz and 1-4kHz). The duplex cues are combined in the IC model, in which it is assumed that the strengths of inputs from MSO and LSO are proportional to the posterior probabilities of the estimated azimuths, that are mathematically calculated by Bayesian theorem. The evaluation dataset is the microphone recorded stereo sounds of five spoken English words. It achieved $90\%$ localization accuracy for sounds between $\pm 45^{\circ}$ with a spatial resolution of $10^{\circ}$, but the performance dropped dramatically when sounds moving to the sides, which means this model can hardly locate the sounds near extreme positions ($\pm 90^{\circ}$). Interestingly, this work is extended to combine with an automatic speech recognition system on a robot system \cite{davila2018enhanced}, to help further improve the speech recognition accuracy. 

Except for directly encoding the duplex cues, some other works offer different views of bio-inspired sound localization mechanisms. Goodman et al. \cite{goodman2010spike} propose that our neuronal system locates the sound source by mapping the binaural location-specific sounds to two synchronous spiking patterns, which are derived from spatio-temporal filtering and spiking nonlinearity, and thus stimulating the activation of location-specific assemblies of postsynaptic IC neurons. In the more recent works, they extend this idea to a learnable, self-adaptive, and noise-robust model  \cite{goodman2010learning}\cite{goodman2013decoding} using Hebbian learning. The experiment results show a mean absolute error of $10^\circ \sim 30^\circ$.








Those aforementioned studies successfully offer novelties in different aspects, such as binaural cues encoding/decoding, or classification back-ends. By summarizing all those works above, we find several common challenges among them, which build a limitation to a more effective and robust neuromorphic approach for real-world sound localization applications:

\begin{enumerate}
    \item Most of the models or architectures are evaluated by the simulated localization datasets. Such datasets are artificially generated by convolving the single-channel audio recordings or pure tones with binaural or multi-channel head-related transfer functions (HRTF) \cite{wall2008spiking,glackin2010spiking,escudero2018real,xiao2016biologically,luke2019spiking}. When applied in real-world recorded sounds, the localization accuracy drops a lot \cite{voutsas2007biologically,goodman2010learning,goodman2013decoding}. 
    
    \item In the application, reliable localization accuracy is only achieved in a small detecting region ($\pm 60^{\circ}$)\cite{liu2010biologically,xiao2016biologically}, or the spatial resolution is not sufficiently high ($20^\circ \sim 30^\circ$)\cite{wall2012spiking,voutsas2007biologically}.
    
    \item Although some works achieve relative good performances, they do not propose purely biologically plausible SNN models with spiking-involved either encoding front-end or decoding back-end \cite{liu2010biologically}\cite{xiao2016biologically}.
    
    \item Most of the models cannot be extended to multiple channels, which is a limitation of the binaural model. Bio-inspired engineering systems always learn from nature and are expected to surpass it.
    
    
    
    
\end{enumerate}


As such, our main contribution in this paper is that we propose a computational model with MTPC front-end and SNN-based learning architecture, for the precise and robust sound localization application in the real-world environment. We use the recorded sounds from a microphone array of four mics, and encode the time differences from each pair of microphones into the spike rates of an assembly of MSO neurons. Then the encoded spikes are fed into different deep architectures of SNNs. By using this model, our system could locate the sound source from $360^\circ$ with a spatial resolution of $5^\circ$. Besides, the MTPC is independent of the temporal synaptic learning rules, as well as of the SNN architectures. So it is friendly to pipeline with most of the existing SNN classifiers.

The rest of this paper is organized as follows: in Section \ref{sec: phase coding} we propose the MTPC scheme to encode ITD cues into spikes. In Section \ref{sec: multi-layer SNN} the recurrent and convolutional architectures of SNN are introduced to learn the spike patterns and estimate the sound source azimuth. Section \ref{sec: experiment} introduces our integrated SSL computational model, the collected dataset by microphone array, as well as the results in various experimental conditions. Discussion and conclusion are in Section \ref{sec: discussion} and \ref{sec: conclusion}.

\section{Neural Phase Encoding of \\ Acoustic Stereo Information}
\label{sec: phase coding}

\subsection{Neural Mechanisms of Coding Interaural Time Differences: the classical Jeffress Model}

\begin{figure}[htbp]
    \centering
    \includegraphics[width=0.5\columnwidth]{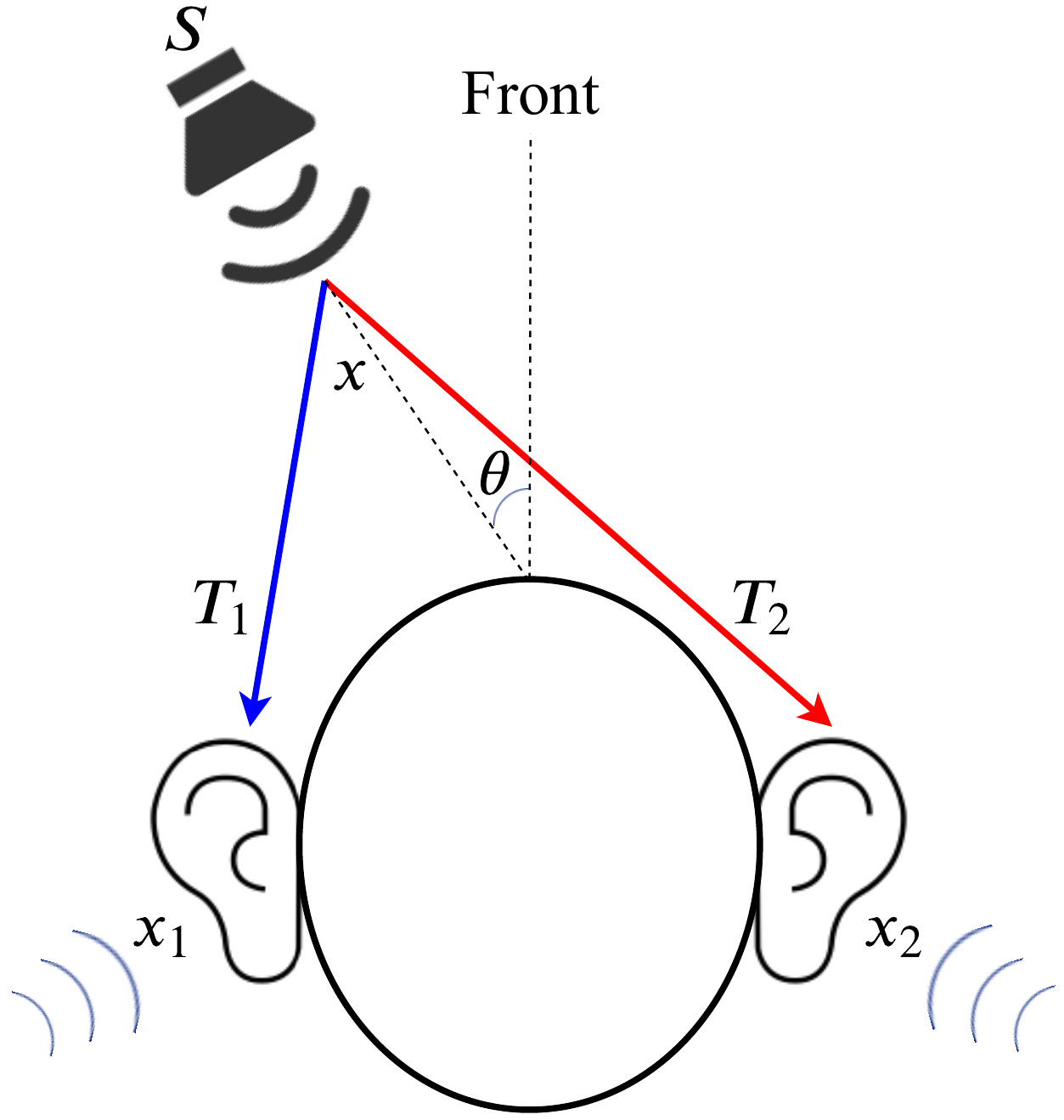}
   \centering
  \caption{Binaural sound source localization problem: a sound source $S$ locates in front of a person's head, $\theta$ degree to the left, and emits a sound signal $x$. $x$ travels through two different pathways (indicated by blue and red arrows) to the left and right ears, becoming $x_1$ and $x_2$, due to suffering different propagation delay, amplitude attenuation, phase distortion, etc. The corresponding propagation times are $T_1$ and $T_2$, respectively. Our goal is to estimate the sound source azimuth $\theta$ from the received distorted sounds $x_1$ and $x_2$, by a spiking neuromorphic approach without any prior knowledge of the transmission channels or other information.}
 \label{fig: problem_description}
\end{figure}

The scene, in which creatures localize the sounds from some particular source by the interaural time difference, can be mathematically formulated as a 2-D sound direction estimation problem, demonstrated in Figure \ref{fig: problem_description}. The sound source emits sound, which travels in the form of mechanical vibration through the air and arrives at two ears separately. The environment includes the air, acoustic noise, obstacles, or acoustic reflectors, that jointly compose the acoustic transmission channel. As such, the received acoustic stimulus by the two ears will suffer different time delay and energy attenuation, which reflects in the changing of phases and amplitudes of the signals.  


\begin{figure}[htbp]
    \centering
    \includegraphics[width=1\columnwidth]{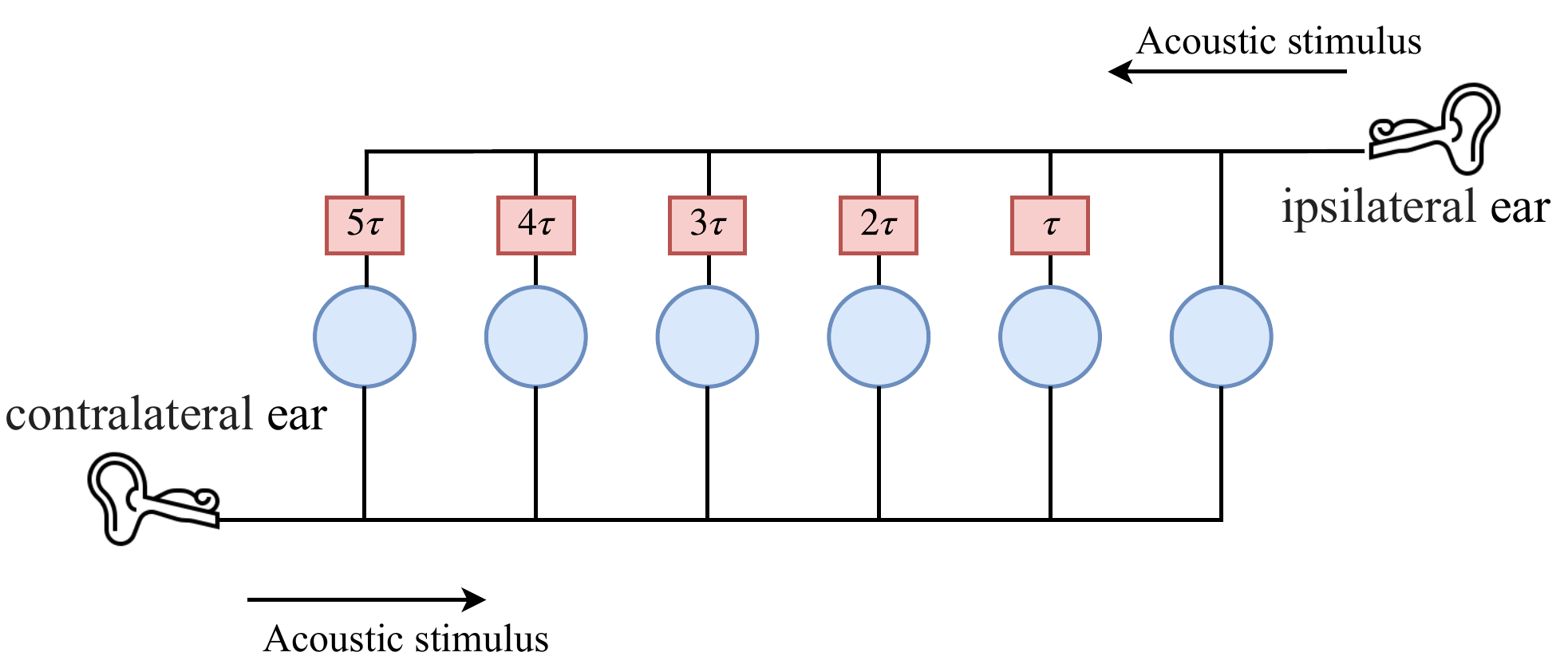}
   \centering
  \caption{Jeffress model of ITD encoding. The sound from a specific location source arrives at two ears separately with different propagation times. The sounds go through each delay line (red boxes), and the blue neuron works as a coincidence detector. For example, let us consider a case where a sound source locates at the left side of $60^\circ$, the sound wave is supposed to arrive at the left ear assuming $3\tau$ earlier than the right ear, where $\tau$ is the smallest processing time unit. As such, the acoustic signal arrived at the right ear will travel through the detector neuron with a synaptic delay of $3\tau$ and meet the one at the left ear simultaneously, which stimulates that neuron to emit significantly more spikes. On the contrary, the detection neurons with other delays will fire little. Such a distribution of firing rates indicates the TDOA between the left and right ears.}
 \label{fig: Jeffress model}
\end{figure}

The sounds from left and right pathways, filtered by the transmission channel, contain the ITD cues implicitly. To estimate the sound source direction, the mammals have evolved an auditory structure known as ``delay line'' in the MSO organ, proposed by Lloyd Jeffress, to encode the mechanical vibrations into electrical pulses. The Jeffress model \cite{jeffress1948place} has become the dominant ITD encoding model since 1948, as demonstrated in Figure \ref{fig: Jeffress model}. The binaural sound waves travel through the pathways of the ipsilateral ear and contralateral ear. Since the sound source occurs at the ipsilateral side, the mechanical vibrations will arrive at the contralateral ear later than the ipsilateral ear. The time difference between the two ears, usually referred to as time difference of arrival (TDOA), is detected by the ``delay line'' structure. The sound at the ipsilateral side ear, referred to as acoustic stimulus in the auditory pathway, passes through a series of neural fiber with different synaptic delays and goes into the coincidence detection neurons in parallel. Once the ipsilateral acoustic stimulus plus some particular delay and the contralateral one arrive at the detection neuron simultaneously, the detection neuron with the particular synaptic delay will fire significantly more spikes. As such, the ``delay line'' model successfully represents the time difference as a firing rate distribution along the detection neurons.

From the computational neuroscience point of view, in principle, the place theory of encoding ITD is to project the time difference onto the spatial domain, which is detected and encoded by a population of neurons. However, this model only offers a prototype of ITD encoding from anatomy. Many models are raised for the application of sound localization, according to the place theory. Among them, however, many problems exist as mentioned in Section \ref{sec: introduction}. So we aim to propose a novel place theory-based neuromorphic computational model that is competitive in the real-world application.

\subsection{Multi-Tones' Phase Coding (MTPC) of ITD localization cues}

\begin{figure*}[htbp]
    \centering
    \includegraphics[width=0.9\textwidth]{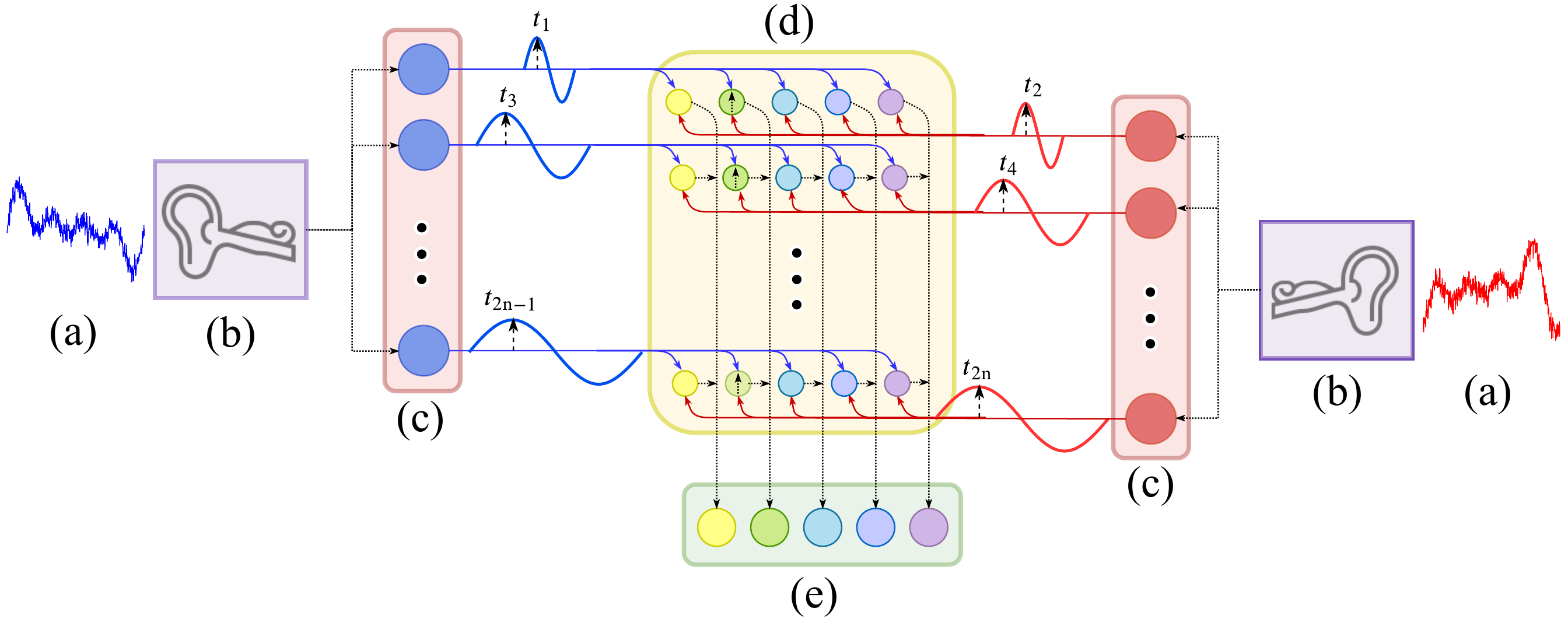}
   \centering
  \caption{Diagram of the three-layer architecture of MTPC for sound localization. The binaural sounds (a) are decomposed into pure tones by frequency analysis in the basilar membrane of the cochlea (b). The spike timings are phase-locked to the first peak of each tone, as indicated by the dash arrows in (c). The phase-locked spike timings are put into the assembling of the coincidence detector neurons (d). For each frequency, only the neuron with a particular synaptic delay that detects the coincidence between the two input spikes will emit a spike to the next layer. In the output neuron layer (e), all the spikes from the coincidence detector neurons belonging to the assembly with the same delay are summed up. The spike rates (number of spikes) constitute the final encoded spike pattern that contains the ITD cues from binaural inputs.}
 \label{fig: whole encoding}
\end{figure*}

Figure \ref{fig: whole encoding} illustrates the MTPC for ITD. This neural spike encoder consists of three main components: the frequency analyzer (b), the phase encoder (c), and the coincidence detector (d). 


\subsubsection{Frequency analyzer}

 Evidence has been provided to prove that our auditory system is primarily a frequency analyzer, in which cells of the basilar membrane are found to respond to different vibration frequencies. Inspired by these acoustic and anatomical findings, our proposed encoder designing begins with a frequency analyzer that decomposes the acoustic input into pure tones with single frequencies, as demonstrated in Figure \ref{fig: first spike phase coding}. In the mathematical implementation, the Fast Fourier Transform (FFT) is used for frequency analysis. Supposing the input sound signal $\bm{x} = [x_1, x_2, ..., x_T]$ with sampling frequency $f_s$, either for left or right ear, is decomposed by $N$-point FFT into an $N$-length complex Fourier series denoted by $\mathscr{F}(\bm{x})$:
\begin{equation}
\begin{aligned}
\mathscr{F}(\bm{x}) &= [a_1+jb_1, ~a_2+jb_2, ~..., ~a_N+jb_N] \\
                    &= [A_1 e^{j\phi_1}, ~A_2 e^{j\phi_2}, ~...,  ~A_n e^{j\phi_n}, \\
                    &\qquad A_{n+1}e^{j\phi_{n+1}},~...., ~A_{2n}e^{j\phi_{2n}}] \\
\end{aligned}
\label{eq: fft}
\end{equation}
Each complex value (i.e. $a_1+jb_1$) from the Fourier series $\mathscr{F}(\bm{x})$ has the physical meaning of the amplitude and initial phase of the corresponding pure tone with a certain analysis frequency. So it is more clear to rewrite them into the form of polar coordinate, such as $A_1 e^{j\phi_1}$, in the second row of Eq.\ref{eq: fft}. Since the FFT results in the analysis bandwidth from 0 to the Nyquist frequency, which is twice higher than the signal's bandwidth, we only need the first half of the $\mathscr{F}(\bm{x})$ vector and ignore the symmetric half. As such, the frequencies of the decomposed pure tones are:  $[\frac{1}{N},\frac{2}{N},...,\frac{n}{N}]*fs$, and $n = N/2$. 

For the $i^{th}~(i=1,2,...,n)$ pure tone, we are able to acquire the analytical expression of the time-domain waveform $\bm{y}_i(t)$ according to the analysis above:
\begin{equation}
\begin{aligned}
    \bm{y}_i(t)&=A_i \text{sin}(2\pi f_i t + \phi_i) \\
    f_i &= \frac{i}{N}f_s\\
    A_i &=\sqrt{a_i^2+b_i^2}\\
    \phi_i &= \text{tan}^{-1}(b_i/a_i)
\end{aligned}
\label{eq: wave1}
\end{equation}
 

\subsubsection{Neural phase encoder}

\begin{figure}[htbp]
    \centering
    \includegraphics[width=1\columnwidth]{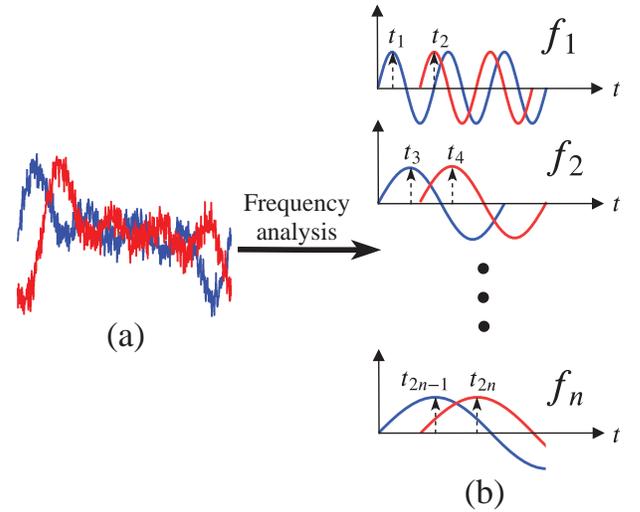}
   \centering
  \caption{Neural phase encoder. The distorted binaural sounds (a) are analyzed in the frequency domain and decomposed into pure tones with frequencies from $f_1$ to $f_n$, as shown in (b). Blue and red curves represent the sounds received by the left and right ear, respectively. As described in Figure \ref{fig: problem_description}, the sound source locates at the left side, so the red curve has some delay to the left curve. Taking the $f_1$ tones as an example, the encoding neurons detect the energy peaks of the oscillation and emit spikes at the first peaks, which are time $t_1$ and $t_2$. Each pure tone has one corresponding phase encoding neuron. The spike timings $t_1$ and $t_2$ embed the ITD cues to be further extracted.}
 \label{fig: first spike phase coding}
\end{figure}

The pure tones, obtained from the decomposing of input sounds by the basilar membrane and frequency-sensitive cells, are encoded into spikes by the auditory nerves in the organ of Corti, which are the electrical pulses transmitted and processed in the spiking neural network. As illustrated in Figure \ref{fig: first spike phase coding}, also in Figure \ref{fig: whole encoding}(c), the spike timings are encoded as the first peak of each analytical pure tones. This is one sort of neural phase coding, which makes the spike timings phase-locked to a fixed phase timing. In this case, the sinusoidal oscillation of acoustic pure tones is phase-encoded by being locked to the peaks. Assuming a pair of pure tones from the left and right ears, $y_1^L(t)$ and $y_1^R(t)$, which has frequency $f_1 = \frac{1}{N}f_s$ and initial phases $ \phi_1^L = \text{tan}^{-1}(b_1^L/a_1^L), \phi_1^R = \text{tan}^{-1}(b_1^R/a_1^R)$, the corresponding phase encoding neurons will fire spikes at $t_1$ and $t_2$:
\begin{equation}
    \begin{aligned}
    t_1 &= (\pi + \text{tan}^{-1}(b_1^L/a_1^L))/2\pi f_1\\
    t_2 &= (\pi + \text{tan}^{-1}(b_1^R/a_1^R))/2\pi f_1\\
    \end{aligned}
\end{equation}

This first-peak phase coding has several advantages. First, we do not need to encode every peak, which is commonly done in phase-locking encoding. It is because the first peaks contain sufficient temporal cues for ITD encoding. Then the first-spike encoding has more energy efficiency than other normal phase codings, the coding efficiency has always been one important aspect of neural encoding front-ends. Lastly, the pure tones are significant, because each pair of pure tones with the same frequency are coherent. From the signal processing point of view, coherent waves are easier for temporal delay detection, signal correlation tasks, compared to complex signals with mixed frequencies. Besides, the phase-locking encoding is also well supported by biological observations.

\subsubsection{Coincidence detector}

\begin{figure}[htbp]
    \centering
    \includegraphics[width=1\columnwidth]{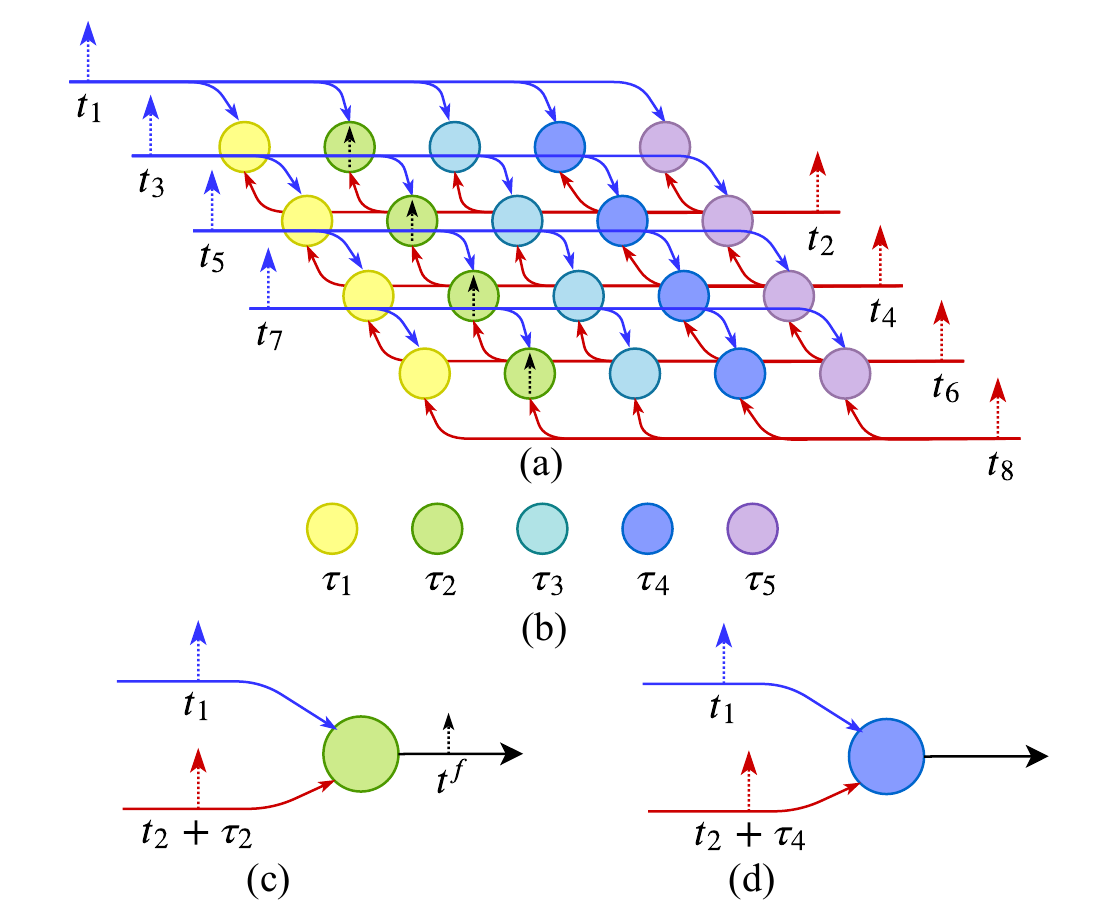}
   \centering
  \caption{Coincidence detection network. The coincidence detection network (a) is composed of different sorts of detection neurons varying in synaptic delays. In this case, five neurons are illustrated in (b). Each sort of neurons have two input afferents to receive the inputs from two ears, shown as the blue and red lines, and one of the afferents has one unique and fixed synaptic delay time. The blue line afferent is designed as a reference line without delay, while the red line has a synaptic delay, varying from $\tau_1$ to $\tau_5$. Each detection neuron will fire a spike if the two spikes arrive at the soma of the neuron simultaneously. For instance, in (c) the spike $t_1$ from the reference ear arrives at time of $t_1$ without delay, while the spike $t_2$ from the other side arrives at time of $t_2+\tau_2$ after a delay of $\tau_2$. Coincidentally they meet at the soma simultaneously with $t_1=t_2+\tau_2$, such that this neuron (c) detects the coincidence and fires a spike. On the other hand, for instance in (d), the neuron will keep silent if the two spikes miss each other with $t_1 \neq t_2+\tau_2$. }
 \label{fig: coincidence detection}
\end{figure}

The last step of the MTPCscheme is the coincidence detection for every pair of phase-coded spike patterns. The detection process also works purely by the transmission of spikes. The schematic diagrams are shown in \ref{fig: whole encoding}(d), and a detailed illustration Figure \ref{fig: coincidence detection}. Coincidence detection aims to extract the ITD information embedded in the phase-coded spikes and represent it in the form of temporal spikes. 

In this process, each pair of phase-coded spikes (refer to Figure \ref{fig: whole encoding}(c)), distinguished by different frequencies, forward into the coincidence detection network (Figure \ref{fig: whole encoding}(d)), and the network outputs spike patterns from Figure \ref{fig: whole encoding}(e).

For instance in Figure \ref{fig: coincidence detection}(a), one pair of binaural location-dependent sounds are decomposed into 4 pairs of pure tones, and they are further phase-encoded into 4 pairs of spike trains: $[t_1,t_2]$, ... ,$[t_7,t_8]$. Assuming that the spike timings with even subscripts are first arriving as in the reference side, the four pairs of spike trains are forwarded into the coincidence detection network. In this network, each row of neurons represents one unique frequency, and each column of neurons with a unified color represents the ITD detection by compensating one unique TDOA. In this noiseless example, the delay of each frequency is all $\tau_2$, such that every row detects the delay of $\tau_2$ and drives the green color neuron firing a spike (black arrows in Figure \ref{fig: coincidence detection}).

The symmetric architecture of the coincidence detection is found in the auditory pathway of mammals, since the compensated delay is naturally a positive real value. However, in the engineering system, negative delay values are implementable. So we apply one coincidence detection network in the engineering realization, instead of the symmetric architecture with two networks, to simplify the computational model and engineering system.

The output spikes of the coincidence detection network organize a 2-D distribution of various frequencies on the pre-defined TDOA range ($\tau_1 \sim \tau_5$). This is a binary matrix in which 1 indicates one spike is fired during the analysis window.  According to the topology of the coincidence detection network (Figure \ref{fig: whole encoding}(d)), the column and row index of each spike on the map figures out the estimation of a particular delay on a certain frequency.




\subsubsection{Output layer of MTPC}

\begin{figure}[htbp]
    \centering
    \includegraphics[width=1\columnwidth]{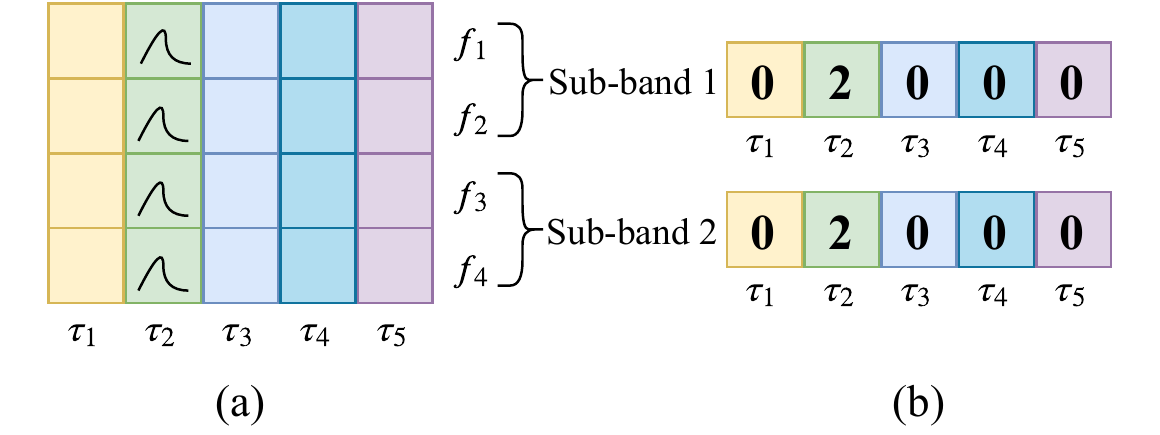}
   \centering
  \caption{The output layer of MTPC. The 2-D output pattern (a) from Figure \ref{fig: whole encoding}(d) will be selectively row-wisely grouped by the frequency channels according to the human cochlear filter bank. The sub-band spike temporal information is represented by the number of spikes emerging within the sub-band. In this illustrative example, four frequencies are grouped into sub-band 1 ($f_1$, $f_2$) and 2 ($f_3$, $f_4$). Within each sub-band, the number of spikes in the two rows is summed up row-wisely. Thus the $4 \times 5$ dimension spike pattern is compressed into $2 \times 5$. In this ideal case, the delays at all pure tones are all precisely detected as $\tau_2$. In a more realistic case, a spike pattern with local peaks and smooth slopes are more common.}
 \label{fig: spike count coding}
\end{figure}

In the final step, the pure tones are grouped into sub-bands in subject to the centre frequencies and bandwidths of human cochlear filter bank, as demonstrated in Figure \ref{fig: spike count coding}. For each sub-band, we sum up the number of spikes along the columns to form a 1-D distribution of estimated delay times. All the 1-D sub-band distribution are stacked over to generate a 2-D pattern (Figure \ref{fig: spike count coding}(b)): one dimension of freqency channel and one dimension of estimated TDOA. 

\begin{figure*}[htbp]
    \centering
    \includegraphics[width=0.9\textwidth]{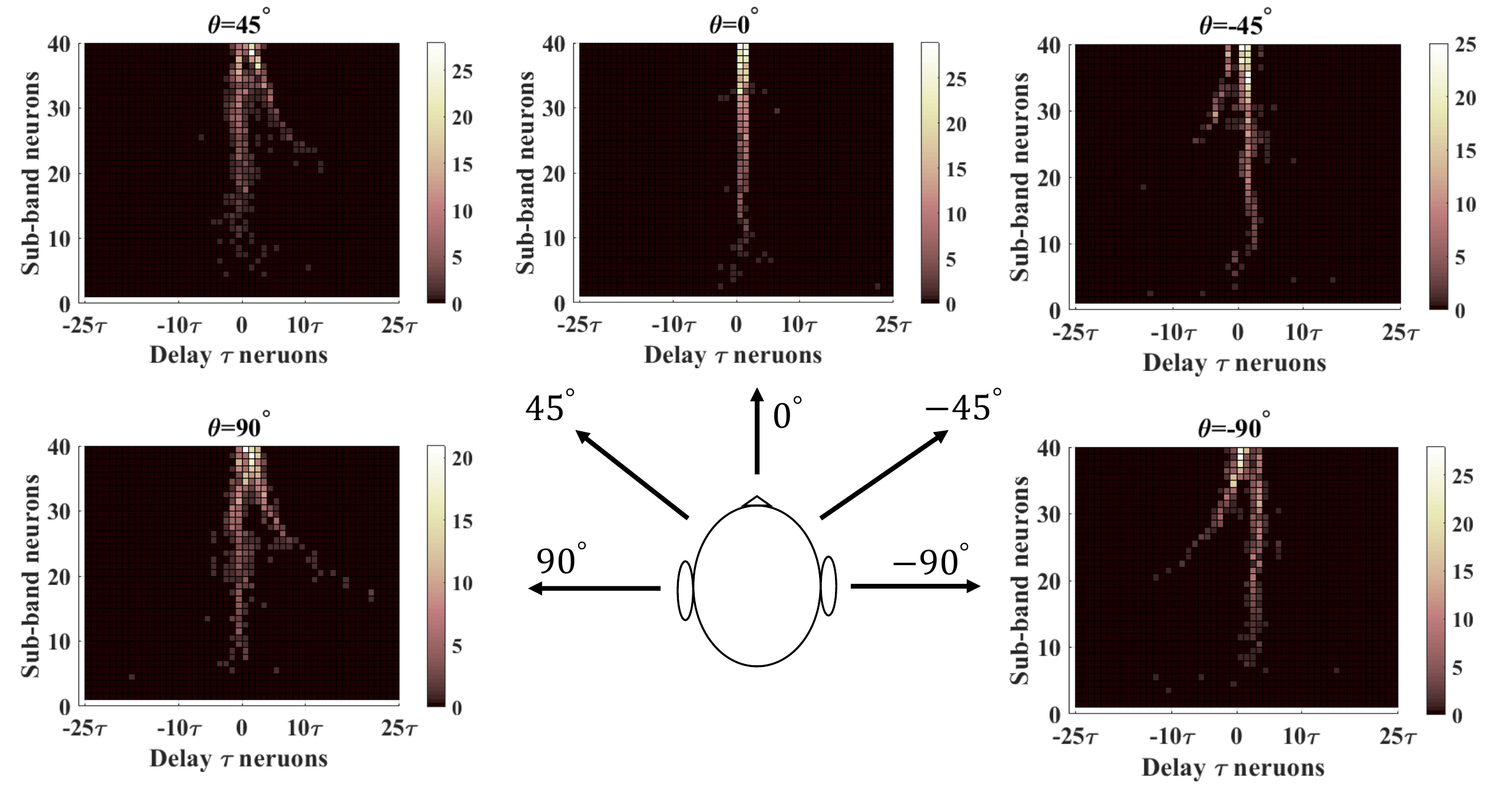}
   \centering
  \caption{Real spike patterns that encode the location-dependent sounds from various azimuths}
 \label{fig: real spike patterns illustraion}
\end{figure*}

Figure \ref{fig: real spike patterns illustraion} illustrates the outputs of the MTPC scheme, by inputting binaural sounds of real-world collected from different azimuths. 40 cochlear channels are utilized in the output layer. The distribution of TDOA on various azimuths is roughly observable. 

\subsection{Biological evidence of the MTPC}

The aforementioned steps are pipelined to generate an integrated neural coding scheme for SSL, the MTPC. In this section, we aim to link the design of MTPC with biological evidence from neuroscience and anatomy.

 
Starting from the binaural case, two ears or microphones receive the acoustic signals $\bm{x}_1$ and $\bm{x}_2$ from two pathways, as demonstrated in Figure \ref{fig: problem_description}. After applying frequency analyzer in Eq. \ref{eq: fft}, each sensed sound wave is decomposed into pure tones. Such a process imitates the tonotopic organization of the cochlea \cite{bourk1981tonotopic}. The basilar membrane, a stiff structural element within the cochlea \cite{ruggero1992responses}, can respond to a certain frequency by some area along with it \cite{ruggero1997basilar}. Such a distribution of single frequencies to various locations along the cochlea is called the tonotopic organization \cite{holton1983micromechanical}. We note that the key to distinguishing the TDOA lies in the spatial variation of frequency responses in the tonotopic organization.

\begin{figure}[htbp]
    \centering
    \includegraphics[width=0.8\columnwidth]{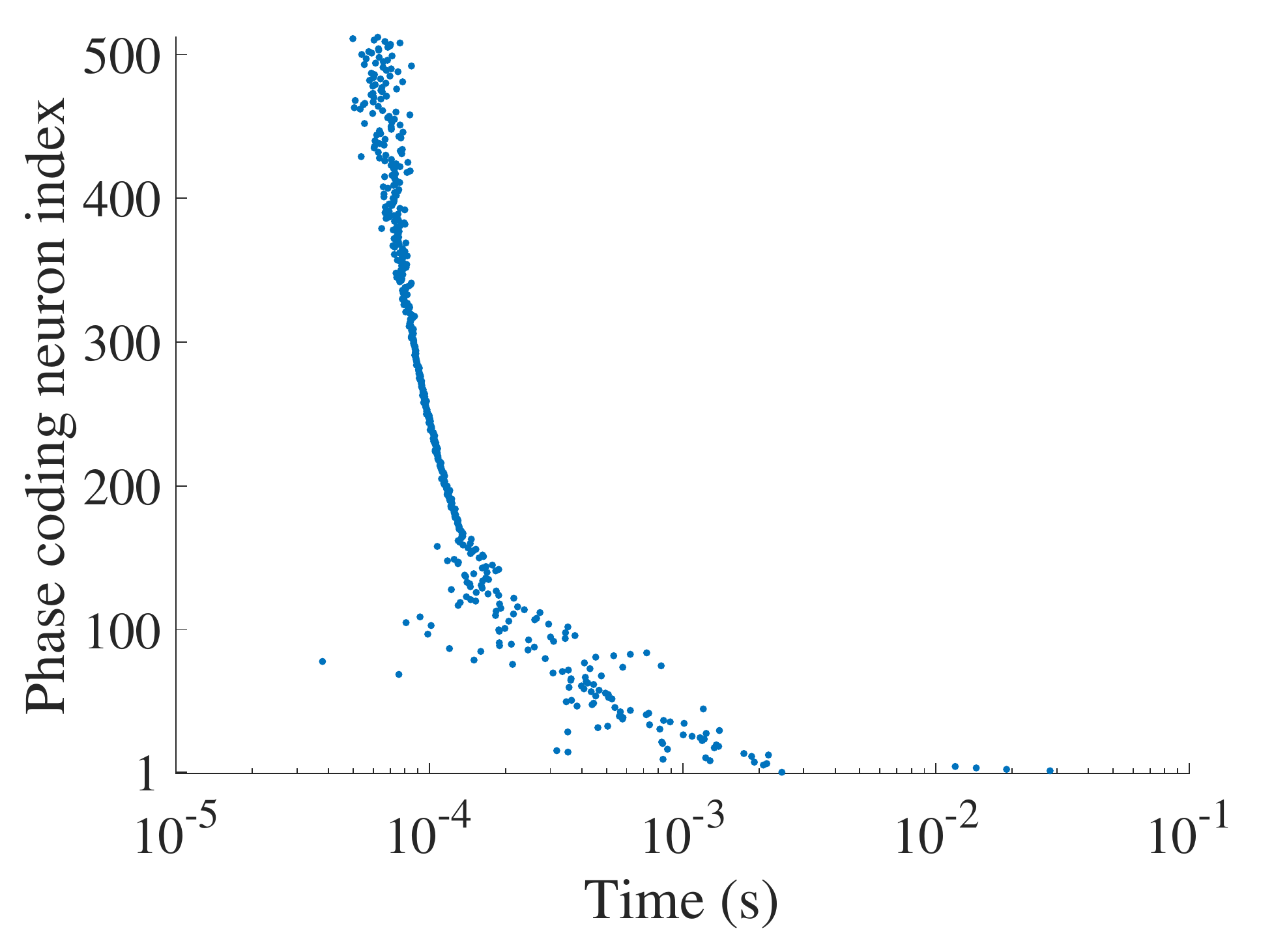}
   \centering
  \caption{Output of neural phase coding. The input sound is decomposed into 512 pure tones, each tone is phase-coded by one neuron. The dots in this figure illustrate the firing timings of each encoding neuron.}
 \label{fig: phase coding pattern}
\end{figure}

The next step is neural phase coding in Figure \ref{fig: first spike phase coding}. The first peak of each tone is detected by the hair cells \cite{brownell1985evoked} in the cochlear and is encoded into a spatio-temporal pattern as in Figure \ref{fig: phase coding pattern}, in which temporal axis represents the peaks' timings and the spatial axis represents the responding locations of the tonotopic organization, that is, the frequencies.

After the neural phase coding, the couple of coded patterns from two ears are fed into the coincidence detection network (Figure \ref{fig: coincidence detection}). In this organization, pair-to-pair comparison of each frequency is conducted, and the frequencies are still distinguished by the spatial variation. This organization is functionally inspired by the Jeffress's finding of delay lines \cite{jeffress1948place}, which anatomically locates at the medial superior olive (MSO) on the mammal's auditory pathway \cite{yin1990interaural}. Finaly the output layer of MTPC connects to the inferior colliculus (IC) \cite{faye1985anatomy} where the localization cues are decoded \cite{winer1998auditory}. In this work, the function of IC is conducted by the spiking neural networks.

\subsection{MTPC for a mixture of single tones: an instance for illustration}

\begin{figure*}[htbp]
    \centering
    \includegraphics[width=0.9\textwidth]{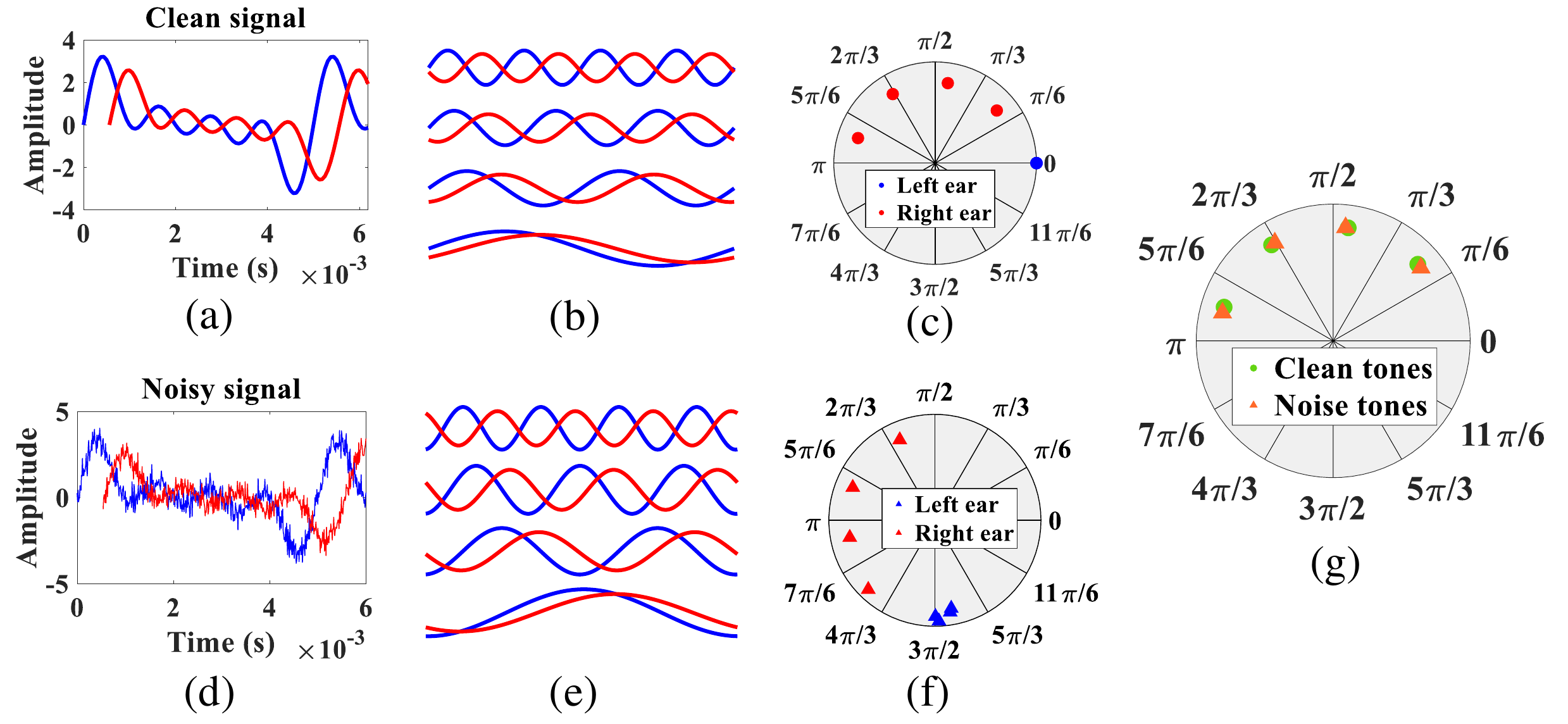}
   \centering
  \caption{The illustration of the phase differences of pure-tone signals. A sound source generates a composed sound with four frequency components of 200Hz, 400Hz, 600Hz, 800Hz. The ideal clean case and real-world noisy case are (a) and (d). By the MTPC scheme, the noisy case can still maintain the phase-differences, from which the TDOA will be directly obtained.}
 \label{fig: Sinewave demonstration}
\end{figure*}

To better understand the mechanism of MTPC in both clean and noisy environments, we illustrate an instance in Figure \ref{fig: Sinewave demonstration}, in which the stereo sounds are composed of four frequencies: 200Hz, 400Hz, 600Hz, and 800Hz. Each frequency tone shares the same energy level. In Figure \ref{fig: Sinewave demonstration}(a) and (d), the binaural stereo sounds are received with a certain TDOA, in clean or noisy environments. Figure \ref{fig: Sinewave demonstration}(b)(e) demonstrate the frequency analysis by the tonotopic organization of the cochlea, that is, the four frequency components are isolated in the time domain by stimulating the vibration of various areas in the basilar membrane. By observation, it is noted that for each frequency, the tones in (b) and (e) begin with different initial phases. It can be expected because the disturbance of white noise makes the phases shift on the sounds. However, what the MTPC scheme cares about is the phase difference between each binaural coupled tones with the same frequencies. Figure \ref{fig: Sinewave demonstration}(c) and (f) shows the initial phases of (b) and (c) in a polar coordinate. Then we obtain the phase-difference for each couple of tones, shown in Figure \ref{fig: Sinewave demonstration}(g), in which the green-colored circles and orange-colored triangles represent the clean and noisy cases. The noisy coupled tones have approximate phase-differences with the clean case, although their phases are shifted by the disturbance of noise. The TDOA of each frequency $f_{tone}$ will be simply derived by:

\begin{equation}
    \delta t_{tone} = \frac{\delta \phi_{tone}}{2 \pi f_{tone}}
\end{equation}
where $t_{tone}$, $\phi_{tone}$, and $f_{tone}$ denote the TODA, phase difference, and frequency of the binaural coupled tones.

In conclusion, by analyzing the phase-differences in single-frequency tones, the TDOA is robust to the disturbance of noise. This point is crucial because for the conventional phase-locking coding strategy, the phase-differences of the original binaural sounds, or sub-band filtered outputs, are so vulnerable to the noise in the real environment that a robust TDOA can hardly be obtained. Such a problem makes the existing phase-lock coding based neural encoding front-ends for SSL only robust to a simple simulated scenario, rather in a real-world environment.

\section{Spiking Neural Networks for Azimuth Prediction}
\label{sec: multi-layer SNN}

The neural encoding MTPC scheme in Section \ref{sec: phase coding} has successfully encoded the ITD cues into spatio-temporal spike patterns. The topology of the spike firings indicates the TDOA that is crucial for azimuth prediction. In this section, we aim to pipeline the encoded firing patterns with different architectures of SNNs, which serve as pattern classification models, to learn the ITD spiking representations and decode the sound source azimuths. To evaluate the universal applicability of the coding scheme, two diverse SNN strategies are exploited. 

\subsubsection{Recurrent connections of LIF  spiking neurons using back-propagation through time}

We first evaluate the encoded pattern in a recurrent connection of spiking neurons, the Recurrent Spiking Neural Network (RSNN). The recurrent connections of neurons are welled approved to be prominent in dealing with temporal dynamic tasks. To optimize the Recurrent Neural Network (RNN), a commonly adopted synaptic updating approach is the back-propagation through time (BPTT).  

The spiking neuron's internal dynamics of sub-threshold membrane potential $V(t)$ is modelled by the leaky integrate-and-fire (LIF) spiking neuron \cite{brette2005adaptive}.
\begin{equation}
    {\tau}_m \frac{dV(t)}{dt} = -V(t) + I(t)
    \label{eq: ch6 LIF model}
\end{equation}

where $\tau_m$ denotes the membrane constant. $I(t)$ represents the synaptic currents from both external input spikes and internal recurrent injection:
\begin{equation}
    I(t) = \sum_i \omega_i^\text{ext} S^\text{ext}(t) + \omega^\text{rec} S^\text{rec}(t)
    \label{eq: ch6 LIF synaptic currents}
\end{equation}
where $\omega$ and $S$ represent the synaptic weights and input spike train, with superscript ext for external input and rec for internal recurrent injection. Once the membrane potential $V(t)$ exceeds the firing threshold $\theta$, the spiking neuron will fire a spike. After that, the membrane potential is reset to 0, and the neuron enters a refractory period during which it cannot spike.


When we organize the aforementioned spiking neurons in a recurrent connection, the Recurrent Spiking Neural Network (RSNN) is defined \cite{bellec2018long}. The synaptic updating rule, which is the backpropagation in recurrent LIF neurons through time, is slightly different from conventional BPTT because the output spike train $z(t)$ of LIF neurons are non-differential. Therefore, researchers propose \cite{courbariaux2016binarized} the pseudo-derivative as a solution and it is proved effective \cite{esser2016cover,bellec2018long}.

\begin{equation}
    \frac{dz(t)}{dV(t)} = \max \left\{ 0, ~ 1-\bigg|\frac{V(t)-\theta}{\theta}\bigg| \right\}
    \label{eq: ch6 pseudo deriv BPTT}
\end{equation}
where $z(t)$ and $V(t)$ denote the output spike train and membrane potential of an LIF spiking neuron.

\subsubsection{Convolutional spiking neural network (CSNN) using spatio-temporal backpropagation}

We also utilize the convolutional connections of LIF spiking neurons (Eq.\ref{eq: ch6 LIF model}-\ref{eq: ch6 LIF synaptic currents}), the CSNN, to evaluate our MTPC scheme. The synaptic updating rule of SCNN follows \cite{wu2019direct}, in which the pseudo-derivative Eq.\ref{eq: ch6 pseudo deriv BPTT} is also used in the propagation of gradients.

\section{Experiment}
\label{sec: experiment}

One of the main superiority of MTPC over the conventional neuromorphic models  \cite{wall2008spiking,glackin2010spiking,escudero2018real,xiao2016biologically,luke2019spiking,voutsas2007biologically,goodman2010learning,goodman2013decoding}, is that the MTPC is designed with robustness to noise, as illustrated in Figure \ref{fig: Sinewave demonstration}. For evaluating the effectiveness in a noisy real-world environment, we have collected our localization dataset using a microphone array; The experiment results evaluate the MTPC scheme in a challenging real-world environment.

\subsection{Real-world collected dataset for precise SSL}
\label{sec: collected dataset}

\begin{figure}[htbp]
    \centering
    \includegraphics[width=0.9\columnwidth]{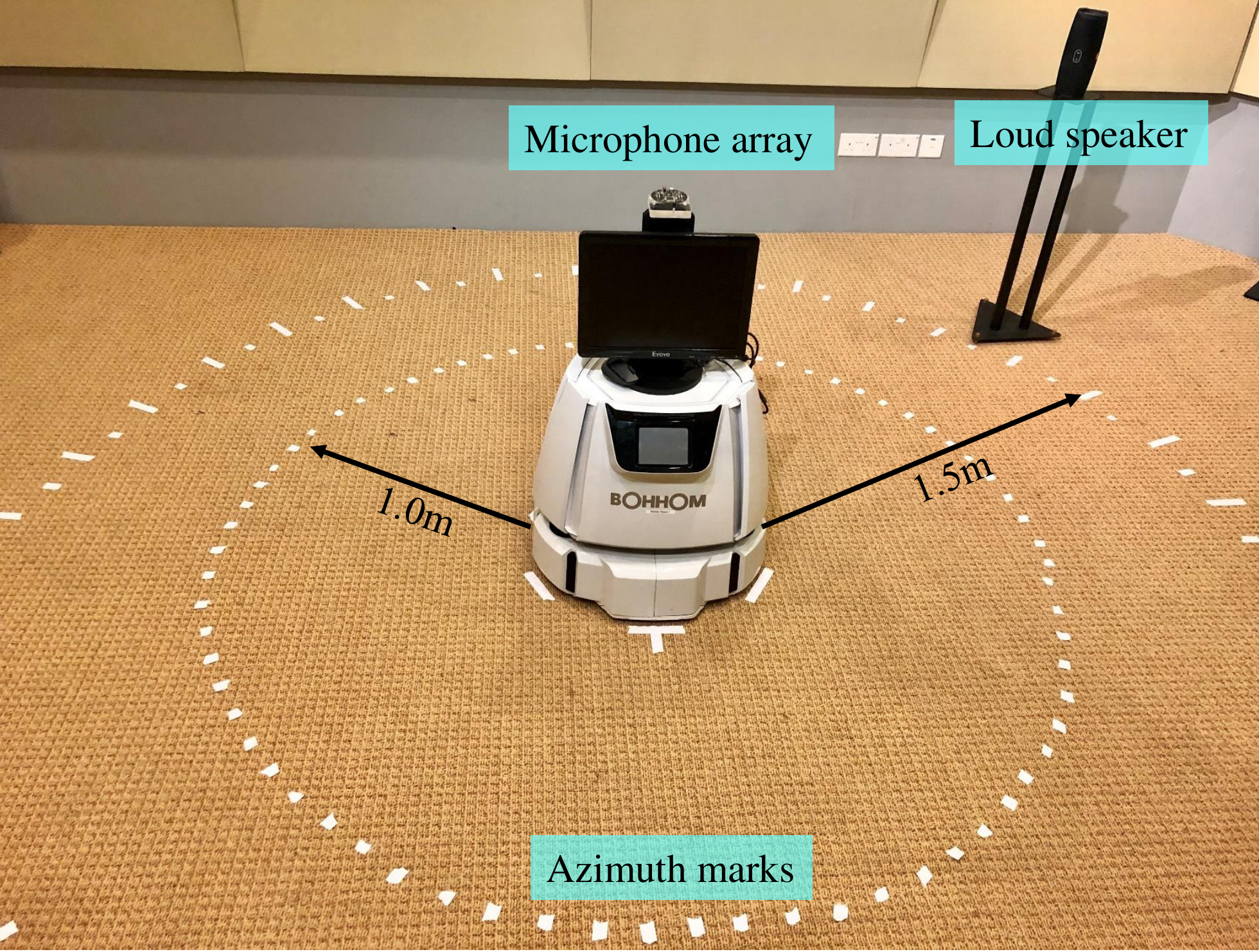}
   \centering
  \caption{The equipment's set-up for the SSL data collection.}
 \label{fig: data collection setup}
\end{figure}

The data collection equipment's set-up is illustrated in Figure \ref{fig: data collection setup}. A microphone array, which locates on top of a robotic platform, records the sounds from a loudspeaker. The loudspeaker moves along the circle to emit stereo sounds from different azimuths. 

\begin{figure}[htbp]
    \centering
    \includegraphics[width=0.9\columnwidth]{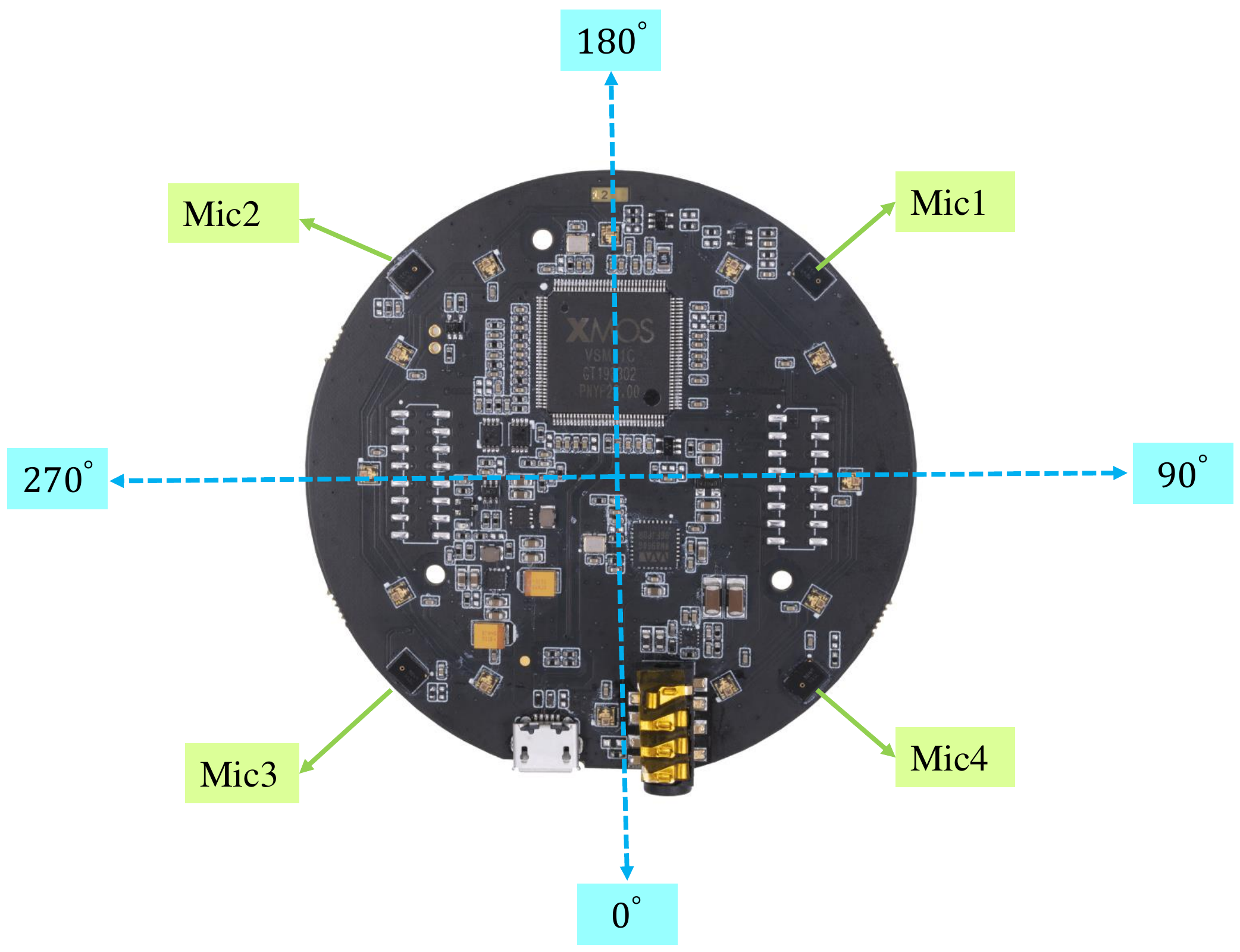}
   \centering
  \caption{The ReSpeaker Microphone array and its settings for the data collection experiment. }
 \label{fig: microphone array}
\end{figure}

The ReSpeaker microphone array is applied in the SSL data collection and its settings are illustrated in Figure \ref{fig: microphone array}. Four microphones, automatically synchronized, record the sounds at 16kHz sampling rate from the loudspeaker and save the waveforms into .wav format files. The $0^\circ$ is defined at the center of Mic 3 and Mic 4, and it increases in counterclockwise to $355^\circ$. The contents of sounds are speech voice from RSR2015 corpus \cite{larcher2014text}, by which we aim to imitate the oral orders to the robot.

In the data collection phase, the multi-functional robot platform serves as a power supplier and support to the ReSpeaker, which stands on the center of the robot. The azimuths are marked on the ground with two distances: 1.0m and 1.5m, as the two circles shown in Figure \ref{fig: data collection setup}. It is also noted that the center of the microphone array overlaps at the center of the marked circles.  

For a constant SNR, the volume of the loudspeaker at 1.5m is set to the maximal value, while it is $\frac{1^2}{1.5^2} \approx 44\%$ of the maximum at 1.0m, according to the inverse square law of wave propagation.

\begin{table}[htbp]
\centering
\begin{tabular}{c|c}
\hline
\textbf{Parameters} & \textbf{Values} \\ \hline
Sampling frequency & 16kHz\\
Num of mic channels & 4 \\
Distance & 1.0m, 1.5m \\
Speech corpus & RSR2015 \cite{larcher2014text} \\

Azimuth range & $[0^\circ,355^\circ]$ \\
Azimuth resolution & $5^\circ$ \\

Total recording time & 15h \\
Sample duration & 170ms\\
Ave training samples/$\text{degree}^\circ$ & 1848 \\
Ave testing samples/$\text{degree}^\circ$ & 344

\end{tabular}%
\caption{The numerical details of the collected SSL dataset.}
\label{tab: numerical details of dataset}
\end{table}

The numerical details of the collected data are shown in Table \ref{tab: numerical details of dataset}. Approximately a total of 15 hours of recordings are collected, and they are clipped into 170ms per sample, with a stride size of 85ms. For each azimuth, the average numbers of training and testing samples are 1848 and 344. The azimuth resolution is $5^\circ$, which is smaller than any known neuromorphic SSL system \cite{wall2008spiking,glackin2010spiking,escudero2018real,xiao2016biologically,luke2019spiking,voutsas2007biologically,goodman2010learning,goodman2013decoding}, which also challenges our proposed model.

\subsection{Integrated spiking-based SSL system for real-world applications}

\subsubsection{Extend from two ears to a microphone array}

The MTPC is inspired by the mammal's binaural auditory pathway for SSL, and explained in the binaural case, though,  its mathematical principle makes it easily extended to a microphone array case, in which topological multiple microphones ($>2$) receive stereo sounds synchronously. 

In the experimental microphone array case, we can regard each pair of mics as one binaural case, and combine them as one individual spike pattern. The reasons are that different mic pairs provide various TDOA features, due to various locations relative to the sound source; and the combination of these features will contribute to estimate the sound source in $360^\circ$ azimuth. 

\begin{table}[htbp]
\centering
\begin{tabular}{c|c}
\hline
\textbf{Parameters} & \textbf{Values} \\ \hline

Number of FFT points & 1024 \\
Number of analytical pure tones & 512\\
Cochlear filter channels & 20, 40\\
TDOA resolution & 0.0625ms

\end{tabular}%
\caption{The parameters of MTPC in the SSL experiment.}
\label{tab: MTPC parameters}
\end{table}

\subsubsection{Apply MTPC to various SNN back-ends}

In this 4-mic case in Figure \ref{fig: microphone array}, a total of $C^2_4=6$ pairs of binaural spike patterns will be achieved. How to combine them into one pattern dependents on the architectures of SNN back-ends. Let's take the CSNN and RSNN as an instance. 

\begin{figure*}[htbp]
    \centering
    \includegraphics[width=0.9\textwidth]{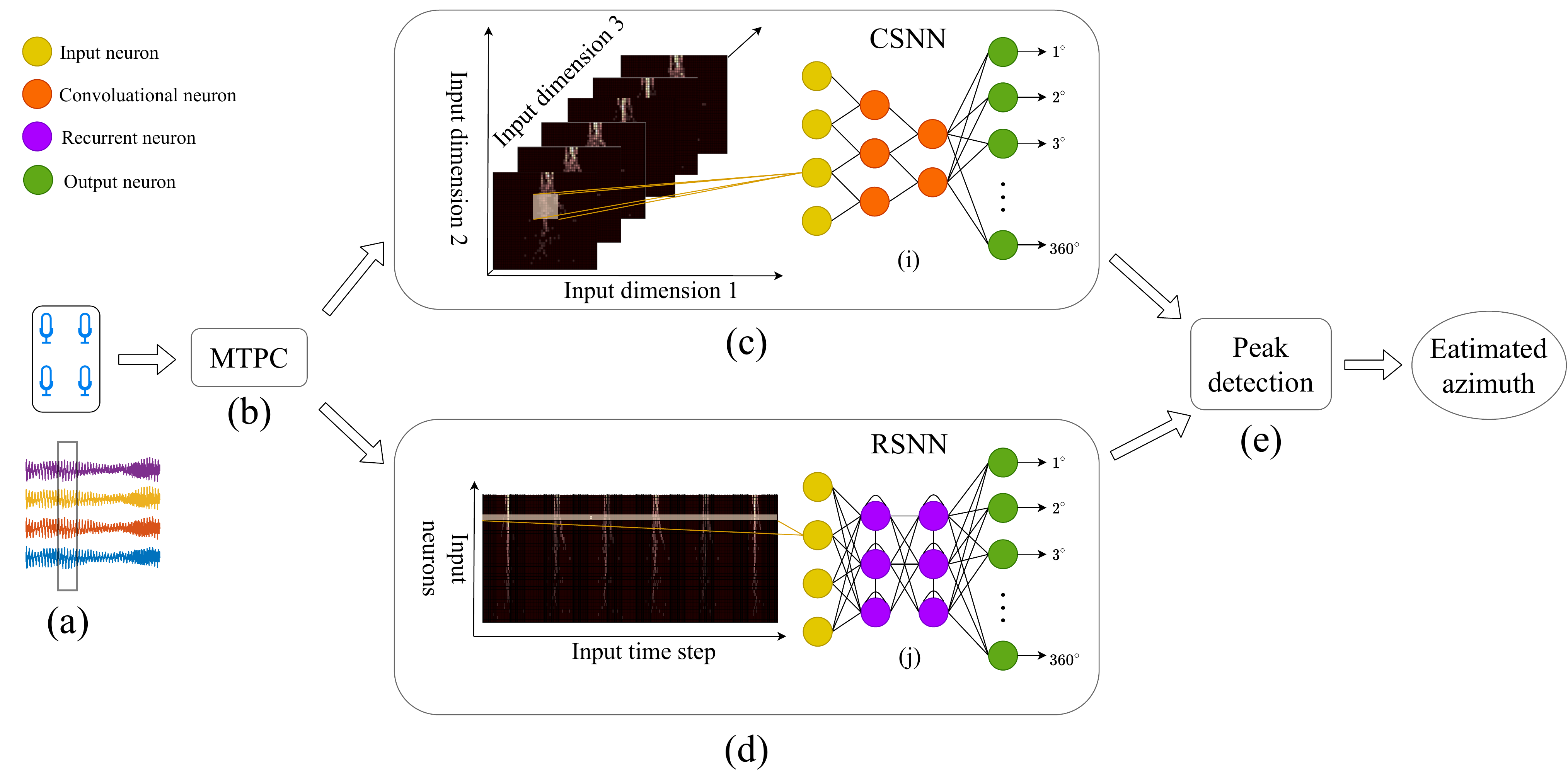}
   \centering
  \caption{Apply MTPC with different SNN structures:the recurrent SNN (RSNN) and convolutional SNN (CSNN)}
 \label{fig: MTPC SNN models}
\end{figure*}

Figure \ref{fig: MTPC SNN models} illustrates the approach to pipeline the MTPC with different SNN architectures. The 4-channel acoustic sample in Figure \ref{fig: MTPC SNN models}(a) is encoded by MTPC into six spike patterns. The six patterns are combined by two approaches:

For the convolutional SNN, the six 2-D spike patterns are stacked into a 3-D spike pattern, shown in Figure
\ref{fig: MTPC SNN models}(c), in which the third dimension is along with the index (1 to 6) of the 2-D patterns. In this case, the encoded neurons are organized in 3 dimensions: the time delay dimension, the frequency channel dimension, and the index of the 2-D patterns dimension.  


For the recurrent SNN, the six patterns are tandemed along the time delay dimension, as in Figure
\ref{fig: MTPC SNN models}(d). In this case, the number of RSNN input neurons are the same as the number of frequency channels. Therefore, each row of the newly combined spike pattern is fed into each input neuron, and the values (the spike counts) are converted into the input current intensities at each time step of the recurrent network.  

The network architectures of CSNN and RSNN are summarized in Table \ref{tab: CSNN and RSNN architecture }. The input layers are corresponding to the dimensions of the MTPC encoded spike patterns; the output layers are both $1 \times 360$ output neurons. The firing rates of the neurons in the output layer indicate the estimated possibilities of sound source's azimuth in each angle around $360^\circ$. The Mean Square Error (MSE) loss function is defined as:
\begin{equation}
    L_{MSE} =  \frac{1}{360}\sum_{i=1}^{360}(\mathcal{L}_{i}-\mathcal{O}_{i})^2
\end{equation}
where $\mathcal{L}$ and $\mathcal{O}$ denote the $1\times 360$ dimensional SNN output and the Gaussian azimuth label curve, respectively. The Gaussian label curve is defined as a $360^\circ$ cyclic normalized Gaussian-shaped curve centered at the azimuth label. At the end, a peak detection block (Figure \ref{fig: MTPC SNN models}(e)) detects the peak of the possibility curve and decides the estimated azimuth in degree. 

\begin{table*}[htbp]
\centering
\begin{tabular}{c|c|c|c}
\hline
\textbf{Spiking neuron connection} & \textbf{Input layer} & \textbf{Hidden layer} & \textbf{Output layer} \\ \hline

CSNN & $51 \times 6 \times 40$ &  12c3s2p1 – 24c3s2p1 - 48c3s2p1 – 96c3s2p1 - fc512  & $1 \times 360$ \\

RSNN & $1 \times 40$ & r1024 & $1\times 360$ \\

\end{tabular}%
\caption{The architecture of SNN back-ends.}
\label{tab: CSNN and RSNN architecture }
\end{table*}. 

\subsection{Localization Accuracy in Different Detection Environments and set-up}

To thoroughly investigate the effectiveness of MTPC, we evaluate the scheme in various experimental environments and compare it with other state-of-the-art SSL algorithms or network models for benchmarking, in terms of Mean Absolute Error (MAE):

\begin{equation}
    \text{MAE} = \frac{1}{N_{\text{sa}}}\sum_{i=1}^{N_{\text{sa}}}|a_\text{Estimate}-a_\text{Lable}|
    \label{eq: MAE equation}
\end{equation}
where $a_\text{Estimate}$, $a_\text{Lable}$, and $N_\text{sa}$ denote the estimated angle and the label angle in degree, and the number of testing samples, respectively.

\subsubsection{Distribution of MAE on different number of coincidence detectors}

In the design of MTPC, we introduce the coincidence detection network (Figure \ref{fig: coincidence detection}) for projecting the ITD of each pure tones onto the delay lines. Given the number of pure tones and frequency channels, the encoded pattern's scale is decided by the number of coincidence detectors, or delay lines. Assuming the number of coincidence detectors is $2d+1$ and sampling rate of $f_s$, the ITD detection ranges in $[-d\frac{1}{f_s}, -(d-1)\frac{1}{f_s},...,-\frac{1}{f_s},0,\frac{1}{f_s},...,d\frac{1}{f_s}]$. Therefore, it is predictable that a wider scale of coincidence detectors will offer a wider range of ITD detection. On the other hand, a larger scale of patterns will bring more computational cost to the SNN and might degrade the performance of it.  

\begin{figure}[htbp]
    \centering
    \includegraphics[width=0.9\columnwidth]{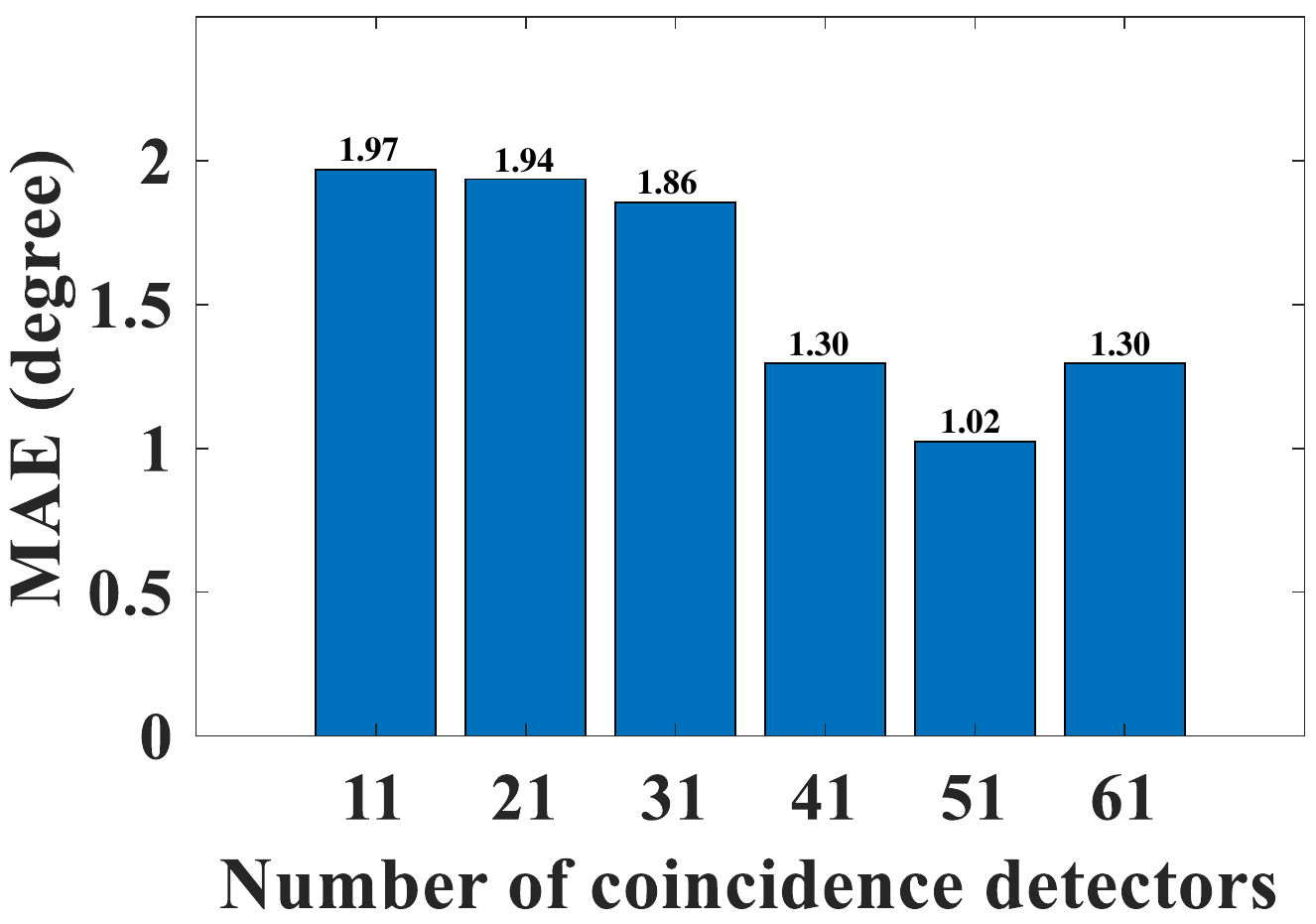}
   \centering
  \caption{The MAE as a function of different numbers of coincidence detection neurons.}
 \label{fig: different tau results}
\end{figure}

We aim to investigate the proper scale of the coincidence detection network, by evaluating the localization accuracy under different numbers of coincidence detectors. Figure \ref{fig: different tau results} demonstrates the MAE as a function of coincidence detector numbers. We test from 11 to 61 coincidence detectors, which means a detection range from $-0.3125\sim 0.3125$ms to $-1.8750ms\sim 1.8750$ms.

We note that the optimal scale of the coincidence detection network is at 51 delay lines, which effectively and efficiently detects the ITD of  $-1.5625ms\sim 1.5625$ms.

\subsubsection{MAE on different frequency channels}

In Figure \ref{fig: spike count coding}, the coincidence-detected pure tones are grouped into frequency channels. We evaluate two cases: 20 channels and 40 channels, for finding the optimal scale of the encoded pattern. Table \ref{tab: 20-40 channels results} shows the SSL accuracy of both cases, by different SNN back-ends. The 40-channel MTPC offers a lower MAE than the 20-channel case. 

\begin{table}[htbp]
\centering
\begin{tabular}{c|c|c}
\hline
\textbf{SNN models} & \textbf{\begin{tabular}[c]{@{}c@{}}$\text{MAE}^\circ$ for \\ 20-channel MTPC\end{tabular}} & \textbf{\begin{tabular}[c]{@{}c@{}}$\text{MAE}^\circ$ for \\ 40-channel MTPC\end{tabular}} \\ \hline
CSNN                & 1.95                                                                                       & 1.61                                                                                       \\
RSNN                & 1.60                                                                                       & 1.02                                                                                      
\end{tabular}
\caption{The MAEs results on 20 and 40 cochlear frequency channels.}
\label{tab: 20-40 channels results}
\end{table}

\subsubsection{Distribution of MAE on $360^\circ$ of all azimuths}

\begin{figure}[htbp]
    \centering
    \includegraphics[width=0.9\columnwidth]{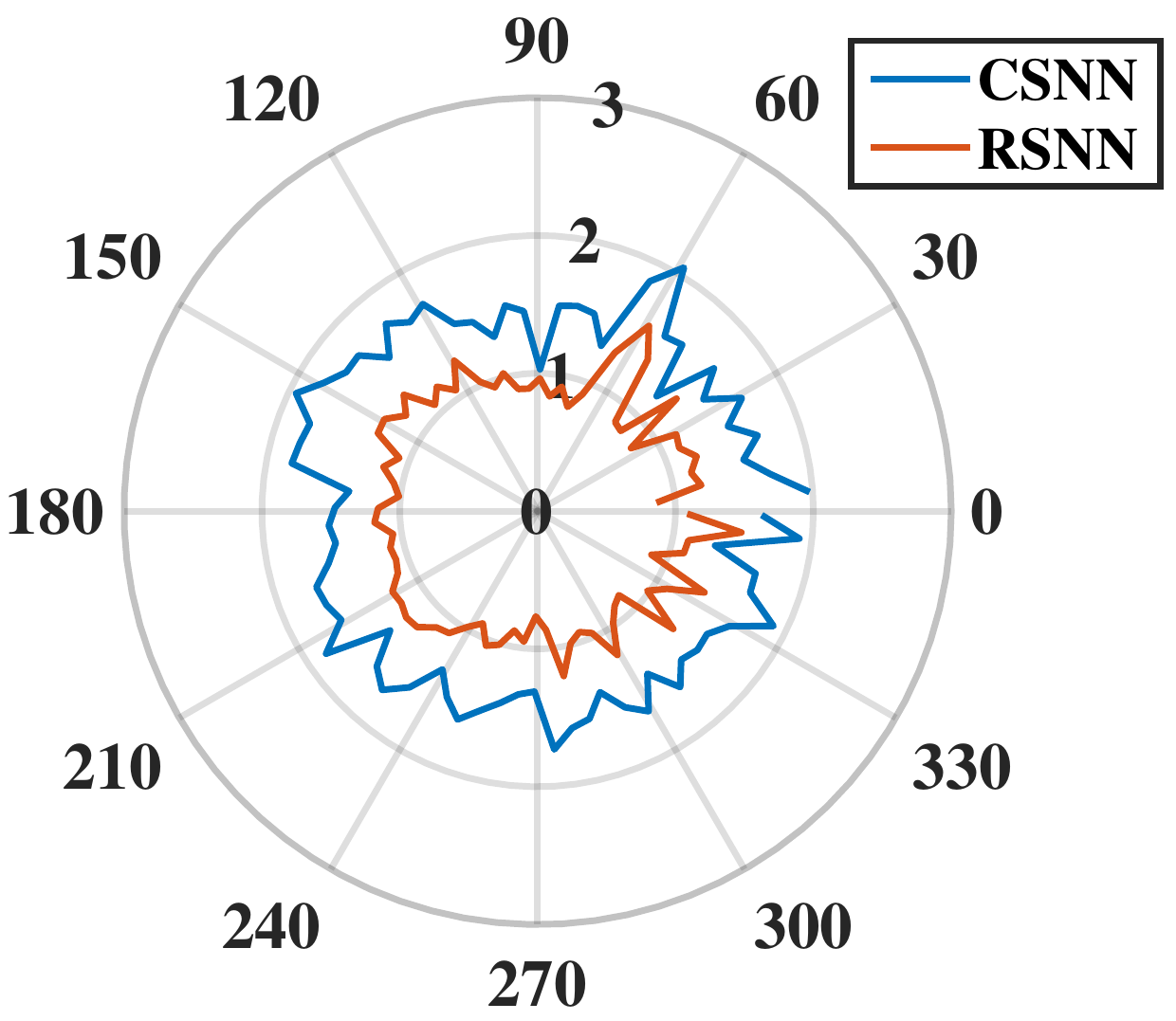}
   \centering
  \caption{The distribution of MAE on  $360^\circ$ all directions, at distance of 1.5m.}
 \label{fig: error distribution}
\end{figure}

A satisfying SSL system is supposed to perform stably on any direction, rather than on some particular ones or ranges. Figure \ref{fig: error distribution} illustrates the MAE distribution on $360^\circ$. At the experimental distance of 1.5m, the results show the MAEs of MTPC with CSNN and RSNN. The MAEs of both models keep lower than $2\circ$ and, more importantly, the MAE performances show no significant difference across all directions. It indicates that: firstly the MTPC scheme achieves stable performance on any direction; and it is extended successfully from binaural to microphone array case.

\subsubsection{Compare the MTPC-SNN models with other state-of-the-art SSL algorithms}

We have already investigated the characteristics of the MTPC itself. Now let's move to compare its performance with the other state-of-the-art  SSL systems.  We make two comparisons:

Firstly, evaluated by the same location-dependent dataset in Section \ref{sec: collected dataset}, the proposed MTPC with various SNN back-ends are compared with two most commonly used non-neuromorphic SSL approaches: the general cross-correlation phase transform (GCC-Phat) \cite{he2018deep} and the multiple signal classification (MUSIC) \cite{schmidt1986multiple} for benchmarking. Experiment results are summarized in Table \ref{tab: SSL compare with GCC-Phat}.

\begin{table}[htbp]
\centering
\begin{tabular}{c|c|c|c}
\hline
Neural Encoding              & Network back-ends     & $\text{MAE}^\circ$ for 1.5m & $\text{MAE}^\circ$ for 1.0m \\ \hline
\multirow{4}{*}{\textbf{MTPC}} & CSNN\cite{wu2019direct}          & 1.61                        & 4.84                        \\  
                               & \textbf{RSNN\cite{bellec2018long}} & \textbf{1.02}               & 4.09                        \\ 
                               & CNN\cite{lecun1995convolutional}           & 1.20                        & 4.06                        \\ 
                               & LSTM\cite{hochreiter1997long}          & 0.41                        & 3.89                        \\ \hline
\multicolumn{2}{c|}{Other SSL Algorithms}      &                             &                             \\ \hline
\multicolumn{2}{c|}{GCC-Phat + CNN\cite{he2018deep}}            & 1.57                        & 4.38                        \\
\multicolumn{2}{c|}{MUSIC\cite{schmidt1986multiple}}                     & 2.35                        & 3.79                       
\end{tabular}
\caption{MAEs results on various SNN back-ends, and comparing with other benchmarking SSL approaches.}
\label{tab: SSL compare with GCC-Phat}
\end{table}

Compared with the other non-neuromorphic algorithms, our MTPC-SNN models outperform them in both 1.5m and 1.0m distances. Our 40-channel MTPC-RSNN models achieve the minimal MAE of $1.02^\circ$. Besides, the MTPC is also pipelined with the conventional ANN back-ends, the convolutional neural network \cite{lecun1995convolutional} and long short term memory (LSTM) network \cite{hochreiter1997long}, making some semi-neuromorphic approaches. It is noted that the MTPC-LSTM performs the best because conventional ANN's ability as classification back-ends is better than that of the SNN, which means proposing more effective learning algorithms for SNN still leaves as an open question.

Secondly, we make a parallel comparison with the other state-of-the-art neuromorphic approaches for SSL. The experimental datasets are simulated or collected by the authors themselves, though, we can compare the complexity and achievements of these works to draw some conclusions. They are summarized in Table \ref{tab: review compare of SSL works}.

\begin{table*}[htbp]
\centering
\resizebox{1\textwidth}{!}{%
\begin{tabular}{c|c|c|c|c|c|c|c}
\hline
\textbf{Proposed by}                                    & \textbf{Spike front-end} & \textbf{SNN back-end} & \textbf{Azimuth range}      & \textbf{Resolution}   & \textbf{Date source}  & \textbf{Sound type}   & \textbf{Results}                                                                  \\ \hline
Xiao, Feng, et al. \cite{xiao2016biologically}      & yes                      & no                    & $-90^\circ \sim 90^\circ$   & $10^\circ$            & HRTF                  & Pure tones            & $74.56\%(\pm 10\circ)$                                                            \\
Wall, Julie A., et al. \cite{wall2012spiking}           & yes                      & yes                   & $-60^\circ \sim 60^\circ$   & $10^\circ$            & HRTF                  & Pure tones            & $95.38\%(\pm 10\circ)$                                                            \\

  Voutsas, Kyriakos, et al. \cite{voutsas2007biologically}   & yes                  & no        & $-45^\circ \sim 45^\circ$    & $15^\circ$           &    Microphone  data     & pure tones          & $72.50\%(\pm 15^\circ)$                                                    \\ 

Liu, Jindong, et al. \cite{liu2010biologically}         & yes                      & no                    & $-45^\circ \sim 45^\circ$   & $10^\circ$            & HRTF                  & Speech                & $90.00\%(\pm 10^\circ)$                                                           \\
Goodman, Dan, et al. \cite{goodman2010spike}     & yes                      & no                    & $-180^\circ \sim 180^\circ$ & $15^\circ$            & HRTF                  & Speech, sounds        & $4^\circ \sim 8^\circ$ MAE                                                        \\
Dávila-Chacón, Jorge, et al.\cite{davila2012biomimetic} & yes                      & no                    & $-90^\circ \sim 90^\circ$   & $15^\circ$            & Microphone  data      & Speech                & $91.00\%(\pm 15^\circ)$                                                           \\
Anumula, Jithendar, et al. \cite{anumula2018event}   &   yes                       & no                     & $-180^\circ \sim 180^\circ$    & $30^\circ$           &    Microphone  data     & Speech           &      $80.00\%(\pm 30^\circ)$                                                                         \\

\multicolumn{1}{l|}{}                                   & \multicolumn{1}{l|}{}    & \multicolumn{1}{l|}{} & \multicolumn{1}{l|}{}       & \multicolumn{1}{l|}{} & \multicolumn{1}{l|}{} & \multicolumn{1}{l|}{} & \multicolumn{1}{l}{}                                                              \\ \hline
40-channel MTPC-RSNN (this work)                        & yes                      & yes                   & $-180^\circ \sim 180^\circ$ & $5^\circ$             & Microphone data       & Speech                & \begin{tabular}[c]{@{}c@{}}$\sim 100\%(\pm 5^\circ)$\\ or $1.02^\circ$ MAE\end{tabular}
\end{tabular}%
}
\caption{Summary of the state-of-the-art neuromorphic approaches for SSL. The comparison is based on the complexities of the datasets and the achievements of the SSL tasks. }
\label{tab: review compare of SSL works}
\end{table*}

We first pay attention to whether their proposed models are purely neuromorphic from spike encoding front-end to SNN back-ends, or semi-neuromorphic with half spiking approaches. It has been mentioned that a purely spike-involved system can fully show all the advantages of neuromorphic computing, such as low power consumption, processing of temporal information, etc. Only 

The complexity of the dataset is evaluated in terms of a larger detection range of azimuth, higher resolution (the minimal distinguished angle), real-world data source other than artificially simulated, and sound types with more frequency components. 

The results are reported in two forms: the correct percentage in some resolution, or MAE. Our 40-channel MTPC-RSNN model outperforms the others and achieves both the highest resolution accuracy and lowest MAE.

\subsubsection{MAE in noisy environments}

Finally, the robustness of the proposed model is examined under various conditions of noisy conditions.

\begin{figure}[htbp]
    \centering
    \includegraphics[width=0.9\columnwidth]{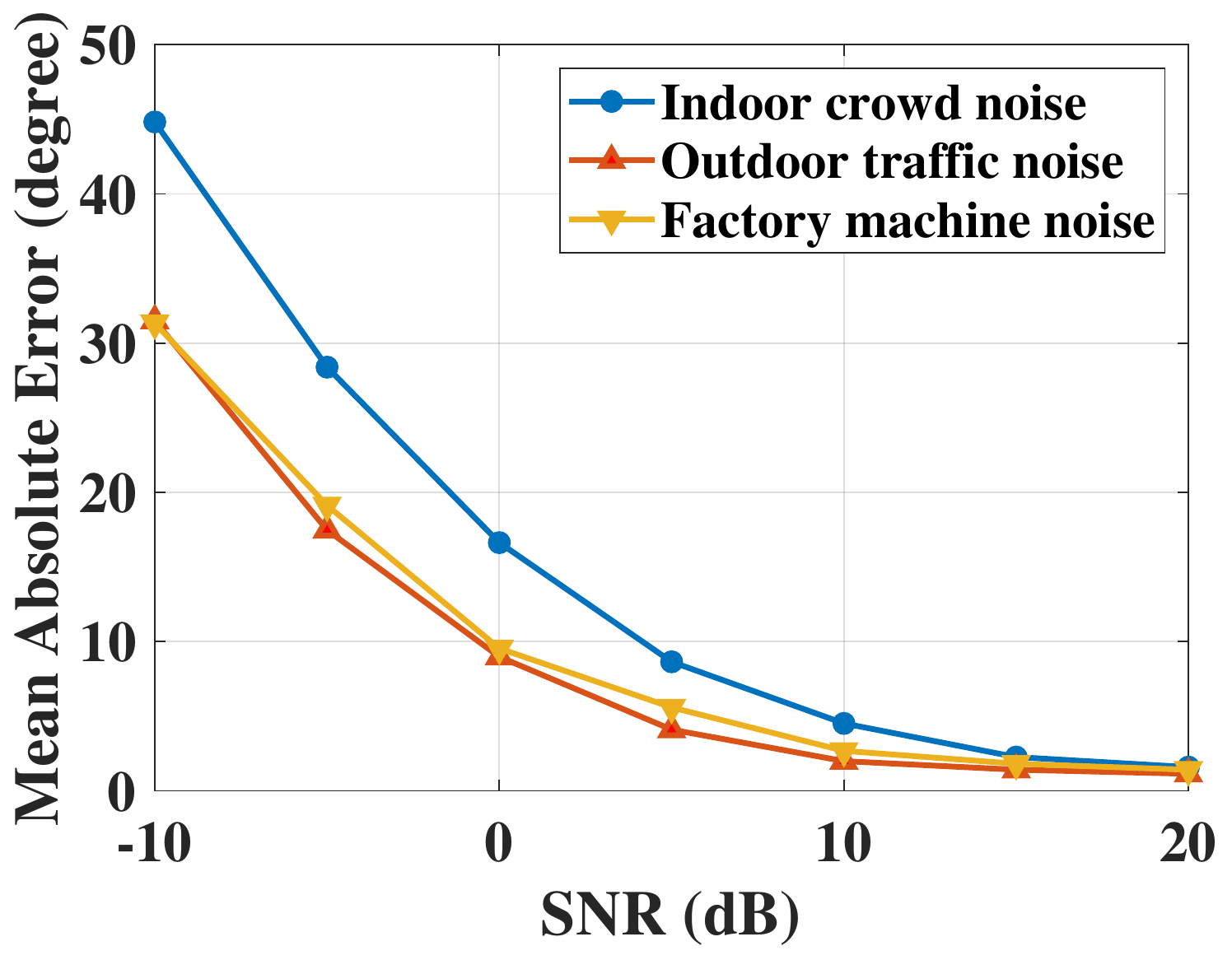}
   \centering
  \caption{The MAE as a function of background noise SNR. The experimental SSL model is 40-channel MTPC-RSNN. The noise robustness is examined under three different types of noise, corresponding to three application scenes in which the SSL systems are usually working.}
 \label{fig: background noise results}
\end{figure}

Figure \ref{fig: background noise results} shows the MAEs as a function of the Signal-to-Noise Ratio (SNR). Three types of background noise: the indoor crowd noise,  factory machine noise, and outdoor traffic noise are used to evaluate the performances in near-field indoor, far-field indoor, and far-field outdoor application scenarios.

Our SSL system performs well (MAE $\le 10^\circ$ ) at SNR$\ge 0$dB. But its accuracy slightly degrades under the indoor crowd noisy environment. This might be because the system is trained with speech-based stereo dataset, thus the system is sensitive to human voices. Furthermore, speech voices have richer and more dynamic frequency components, compared with the traffic or machine noise.

The next experimental scenario is the directional noise of human speech. The sound from the detected source is disturbed by the sounds from other directions. In this experiment, we arrange the disturbance source of speech voice at 1.5m distance, randomly from one of the four directions: $0^\circ$, $90^\circ$, $180^\circ$, $270^\circ$.

The experiment is conducted by conditional training. According to the  SNR, the directional noise are grouped into two levels: high-noise (0-5dB) and low-noise (15-20dB). The conditional training means that the SSL model is trained and tested with noisy data from the same noise condition level. And the clean training means that the model is trained with noiseless data and tested with noisy data. 

\begin{table}[htbp]
\centering
\begin{tabular}{c|c|c|c|c|c}
\hline
\textbf{Testing SNR (dB)} & \textbf{0} & \textbf{5} & \textbf{10} & \textbf{15} & \textbf{20}\\ \hline
Clean training & $54.84$ & $33.93$ & $12.64$ & $8.75$ & $3.88$ \\
Conditional training   & $10.75$ & $6.83$ & $-$ & $1.76$ & $1.07$ \\
\end{tabular}%
\caption{The MAE of conditional training in different directional noise SNR. The acoustic disturbance of human speech randomly comes from one of the four directions: $0^\circ$, $90^\circ$, $180^\circ$, $270^\circ$. }
\label{tab: directional noise results}
\end{table}

This SSL system is also designed to be applied in the robotic platform, that receives the oral orders of human speech. As such, the robustness to the directional human voice noise is significant. The results in Table \ref{tab: directional noise results} indicate the effectiveness of conditional training, in which MAE drops a lot at a high-noise level. Conditional training of the SSL system offers us an effective solution to directional noise.

\section{Discussion}
\label{sec: discussion}

The MTPC scheme is independent of the types of synaptic learning rules, no matter they are rate-based or temporal-based; and also independent of the architectures of SNN, no matter recurrent or convolutional connections. The principle of MTPC is, by following the auditory pathway of mammals, to project the implicit ITD cues onto the time-delay dimension and frequency-channel dimension (Figure \ref{fig: real spike patterns illustraion}), which explicitly demonstrates the distribution of TDOA across various frequency sub-bands. The values of the pattern, interpreted in mathematics, represent the confidence of the particular TDOA. If they are interpreted in a neuronal system, the values could be either spike rates (for rate-based SNN) or synaptic currents (for temporal-based SNN), by which the higher values means the more intensive inputs. Furthermore, as long as the time-delay dimension is preserved, the spiking pattern is compatible with most sorts of SNN architectures, which is proved by CSNN and RSNN in the experiments.


We note that the MAEs for 1.0m are always higher than those for 1.5m (Table \ref{tab: 20-40 channels results}, \ref{tab: SSL compare with GCC-Phat}), which might be counter-intuitive. This is because the loudspeaker occupies some space (Figure \ref{fig: data collection setup}), which covers a range of azimuth when the distance is nearer. Therefore, the data labeling is not accurate, since it is not distinguishable at such a fuzzy range. Nevertheless, considering the detection distance, the MAE of $\sim 5^\circ$ is sufficiently accurate. 
\section{Conclusion}
\label{sec: conclusion}

In this paper, we propose a novel neural encoding scheme: the Multi-Tones' Phase Coding for the SNN-based SSL task. This MTPC is able to encode the input raw waveforms, either from ears or microphone array, into spike patterns that are compatible with most of the existing SNN architectures and synaptic learning methods. The significance of this work is achieving a purely neuromorphic computational model that fully owns the advantages over the ANN models, superior to the other semi-neuromorphic approaches. Moreover, our model outperforms the other SNN-based models, in terms of the dataset complexity, detection range, localization accuracy, etc. 


%



\section*{Acknowledgment}
This research is supported by Programmatic grant no.
A1687b0033 from the Singapore Government’s Research, In-
novation and Enterprise 2020 plan (Advanced Manufacturing
and Engineering domain)

\ifCLASSOPTIONcaptionsoff
  \newpage
\fi

\bibliographystyle{IEEEtran}
\bibliography{IEEEabrv}

\end{document}